\documentclass[fleqn,usenatbib]{mnras}
\usepackage{newtxtext,newtxmath}
\usepackage[T1]{fontenc}
\DeclareRobustCommand{\VAN}[3]{#2}
\let\VANthebibliography\thebibliography
\def\thebibliography{\DeclareRobustCommand{\VAN}[3]{##3}\VANthebibliography}
\usepackage{graphicx}	% Including figure files
\usepackage{subfigure}
\usepackage{hyperref}
\usepackage{xcolor}
\usepackage{lscape} 
\usepackage{multirow}
\title[DL for estimating parameters of GWs]{Deep Learning for estimating parameters of Gravitational Waves}

\author[Shashwat et al.]{
Shashwat Singh$^{1,a}$,\thanks{E-mail: shashwat98singh@gmail.com}
Amitesh Singh$^{1,a}$, %\thanks{E-mail: amiteshsingh487@gmail.com}
Ankul Prajapati$^{2,a}$,
Kamlesh N. Pathak$^{1,a}$
\\
$^{1}$Sardar Vallabhbhai National Institute of Technology, Surat, India- 395007\\
$^{2}$Sorbonne Université, Paris - 75005, France\\
$^{a}$ Bose.X TRIAC Collaboration\thanks{\url{https://www.bosex.org/}}}

\date{Accepted 2021 August 18. Received 2021 August 17; in original form 2020 September 9}

\pubyear{2021}

\begin{document}
\label{firstpage}
\pagerange{\pageref{firstpage}--\pageref{lastpage}}
\maketitle

% Abstract of the paper

\begin{abstract}
In recent years, improvements in Deep Learning (DL) techniques towards Gravitational Wave (GW) astronomy have led to a significant rise in the development of various classification algorithms that have been successfully employed to extract GWs of binary blackhole merger events from noisy time-series data. However, the success of these models is constrained by the length of time-sample and the class of GW source: binary blackhole and neutron star binaries to some extent. In this work, we intended to advance the boundaries of DL techniques using Convolutional Neural Networks, to go beyond binary classification and predict the physical parameters of the events. We aim to propose an alternative method that can be employed for realtime detection and parameter prediction. The DL model we present has been trained on 12s of data to predict the GW source parameters if detected. During training, the maximum accuracy attained was 90.93\%, with a validation accuracy of 89.97\%.
\end{abstract}

\section{\textbf{Introduction}} \label{sec:intro}
Gravitational Waves (GWs) are the disturbances in spacetime generated by highly energetic astrophysical events, high enough to disturb the spacetime. The highly energetic astrophysical events are the source for this radiant energy that propagates in the form of ripples traversing away from the source. GWs can be produced only with astrophysical events involving extreme mass-densities as outlined in the \cite{2} review. Traversing at speeds closer to the relativistic limit, they have a weakly interacting nature with all known matter in Standard Model, further making them challenging to detect. The GWs once merely predictions were strongly confirmed by the LIGO-Virgo Scientific Collaboration (LSC)\footnote{\url{https://www.ligo.org/}} on September 15, 2015 (\cite{GW150914}). Advanced LIGO (aLIG) (\cite{design}) and the Advanced Virgo (aVIRGO) (\cite{acernese2015advanced}) are highly sophisticated, highly enhanced Michelson Interferometers with two arms extending to 4 kilometers for LIGO and 3 kilometers for VIRGO, perpendicular to each other. The detectors measure the \texttt{spacetime strain}: strain induced in spacetime due to distortions by traversing GWs. The GWs from the event disrupts the coherence pattern between the lasers in the arms capturing the strain induced. The interferometer produces this data as a continuous set of strain values in time which is publicly available at \url{https://www.gw-openscience.org/data/}.

The highly sensitive interferometers may detect the signals from other transient sources, such as core-collapse supernovae (CCSNe) (\cite{c2}), gravitational wave orphan memory (\cite{PhysRevLett.118.181103}), cosmic strings (\cite{cs}), or an unknown astrophysical. Being sensitive makes it prone to the most trivial and unpredictable noise disturbances. Sometimes, the noise signals superimpose the real signal, seldom following a specific pattern and repeating themselves during the run. Various signal processing techniques can remove most random non-Gaussian noise, but some noises are insidious and considerably affect the signal either by mimicking or masking a part or whole of the real signal resulting in a false detection. Noise signals that are repeating have a significant characteristic, are transient and non-Gaussian in nature, known colloquially as glitches (\cite{blackburn2008lsc}). Thus, the strain data encompasses information not only of the astrophysical source but also the physical conditions that affected detection quality.

In this work, we aim to present a Deep Convolutional Neural Network (DCNN) model aiming to predict specific parameters of the strain data ($12$\footnote{23 labels count} parameters in total). The parameters we train over are selected in such a way that the model predicts the parameter space of the GW source if detected. The different GW events that we have aimed for detection are Blackhole binaries (BHB), Neutron stars binaries (NSB), Gravitational-wave Echoes (\cite{echo1}), Core-collapse supernovae (CCSNe). The brief detail and waveform constructing of the events have been discussed in Sec-\ref{sec:level2}. 

The structure of this paper is as follows: in Sec-\ref{sec:level2} we briefly describe the GW events that we have aimed for, along with insights on how the noise and glitches affect their detection; their probable sources, and specific features. Sec-\ref{3} focuses on previous contributions towards extracting GW signals' parameters using DL. In Sec-\ref{5}, we discuss the features of the model, methodology of data preparation, model selection, model preparation, and thereon justification for the same and finally testing and validation. Finally, we conclude with the discussion of the results in Sec-\ref{6}.

\section{\textbf{Overview and Components of Strain data}}
\label{sec:level2}
In this section, we have discussed the waveform generation schemes for the GW events used in the dataset. Blackhole binaries (BHB), Neutron stars binaries (NSB), Gravitational-wave Echoes, Core-collapse supernovae (CCSNe), Detector noise, and Glitches have been discussed in this section.

\subsection{\textbf{Black hole and Neutron Star binaries}}
\label{2-A}
Blackhole binaries (BHB) can be studied under three stages of signal generation: inspiral, merger, ringdown. The inspiral part is generated by using Post-Newtonian (PN) approximation solving for the PN parameter $ x = \frac{v^{2}}{c^{2}}$ from Eq-5 in \cite{buskirk2019complete} and the merger part is solved using the semi-analytical method of the original nonlinear equations of General Relativity (\cite{23}). The amplitude of GWs can be constructed as a combination of two independent states of polarizations, namely $h_{\times}$ and $h_{+}$, represented as a function of time $t$, and polar and azimuthal angles $\theta$ and $\varphi$, for the direction of propagation. These dimensionless quantities in the fundamental frame can be evaluated as in Eq-\ref{3a}.
\begin{multline}\label{3a}
h_{+}=-2M\eta\big[(\dot{r}^{2}+r^{2}\dot{\Phi}^{2}+\frac{M}{r})\cos{2\Phi}+2r\dot{r}\dot{\Phi}\sin{2\Phi}\big]\\
h_{\times}=-2M\eta\big[(\dot{r}^{2}+r^{2}\dot{\Phi}^{2}+\frac{M}{r})\sin{2\Phi}+2r\dot{r}\dot{\Phi}\cos{2\Phi}\big]\\
h(t,\theta,\varphi;\lambda)= h_{+}(t,\theta,\varphi;\lambda)+ih_{\times}(t,\theta,\varphi;\lambda) \\
\end{multline}

Development in Numerical Relativity (NR) led to the construction of NR-PN hybrid waveform template-banks. These surrogate waveforms require a reduced parameter space to generate the waveform  (\cite{26}, \cite{27}). Since its development, plenty of codes are available that are capable of accurately and precisely simulating all the phases of a BHB-system: inspiral, merger, and ringdown, thereby extending the results towards extreme spins and high mass ratios (\citep{37}, \citep{34}, \citep{38}). For dataset generation scheme we used PyCBC\footnote{\url{https://www.gw-openscience.org/data/}} pipeline that enables the use of LALSuite simulations\footnote{\url{https://git.ligo.org/lscsoft/lalsuite}}. In Sec-\ref{5-A} we have summarized the PyCBC pipeline and highlighted the sample data generated for the model using the same pipeline.

For Neutron Star binaries (NSB), an in-depth overview of numerical treatment can be developed from reviews  \cite{nsb1} and \cite{nsb3}.

\subsection{\textbf{Gravitational Wave Echoes}}
\label{2-D}
\begin{figure*}
\begin{subfigure}[CIE waveform for $\tilde N_{echo}$ = 7]{\includegraphics[width=7.5cm,height=5.5cm]{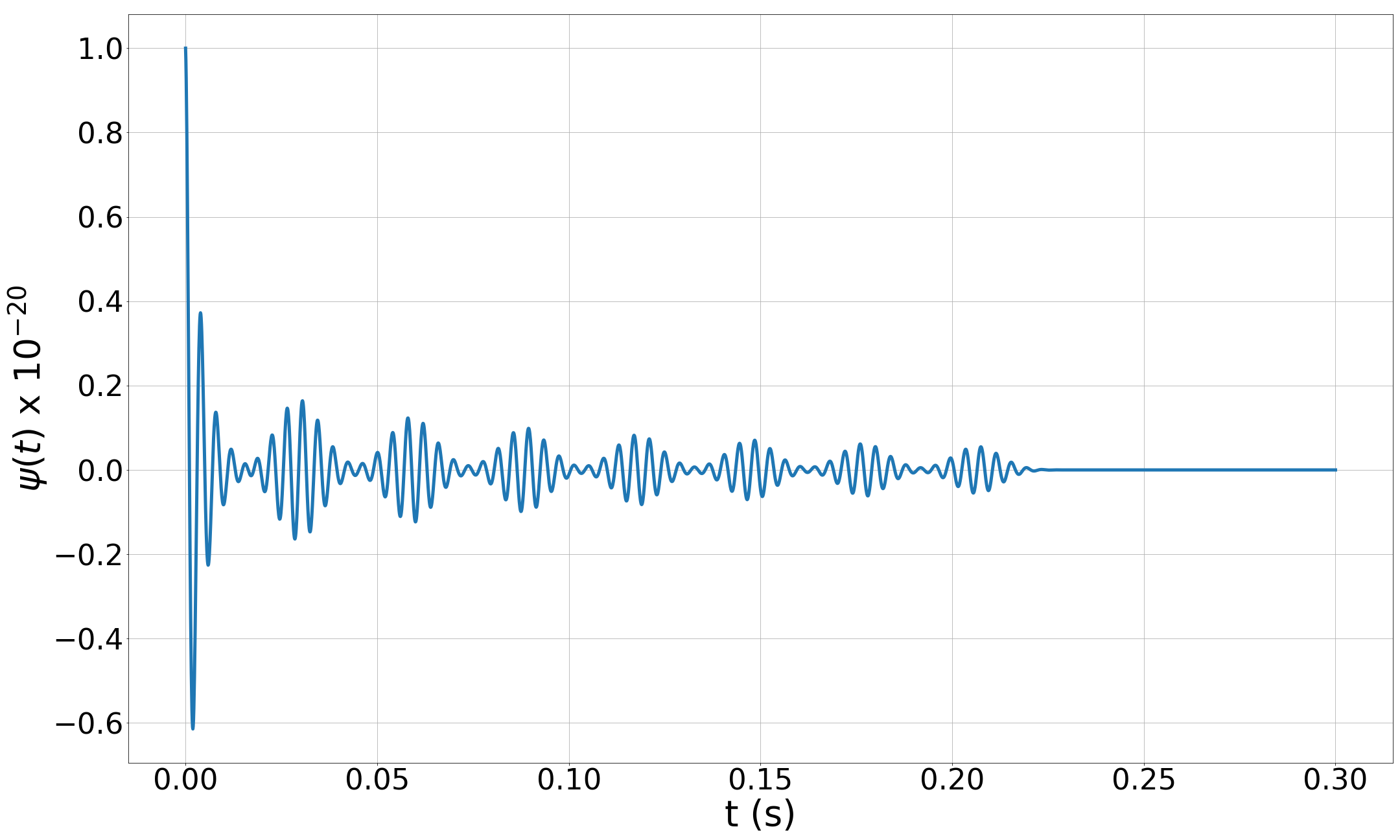}}\end{subfigure}
\begin{subfigure}[CIE waveform for $\tilde N_{echo}$ = 9]{\includegraphics[width=7.5cm,height=5.5cm]{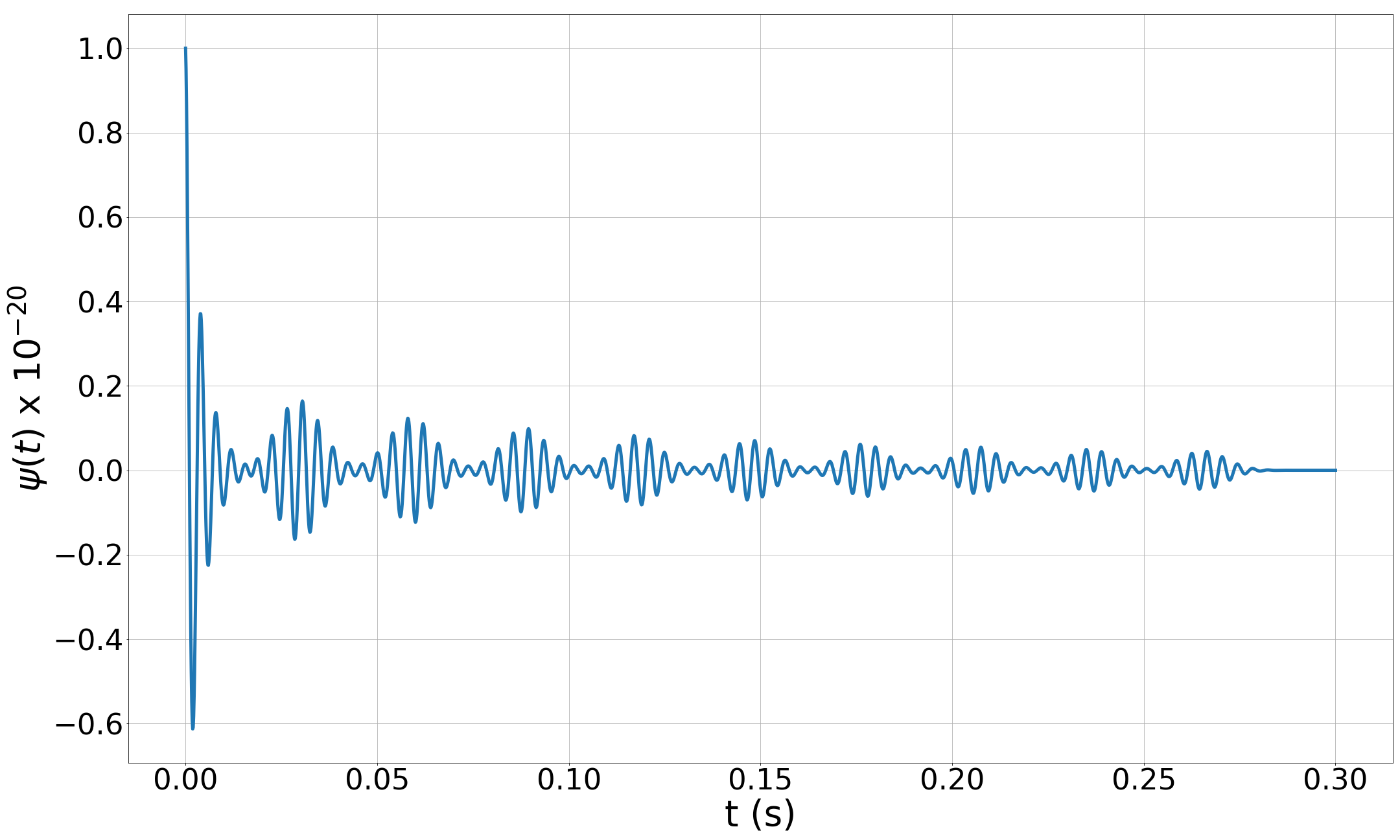}}\end{subfigure}
\begin{subfigure}[UIE waveform for $\tilde N_{echo}$ = 3]{\includegraphics[width=7.5cm,height=5.5cm]{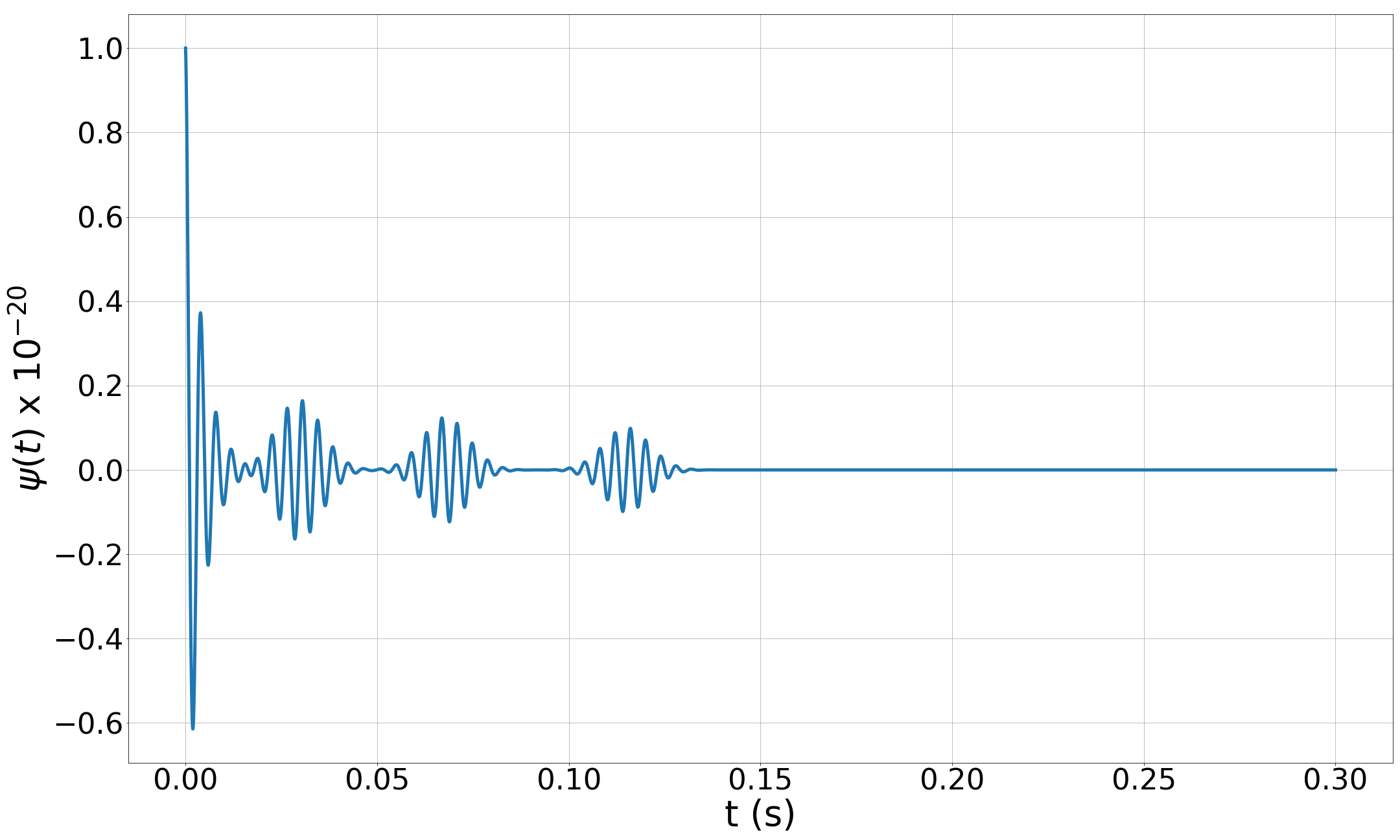}}\end{subfigure}
\begin{subfigure}[UIE waveform for $\tilde N_{echo}$ = 5]{\includegraphics[width=7.5cm,height=5.5cm]{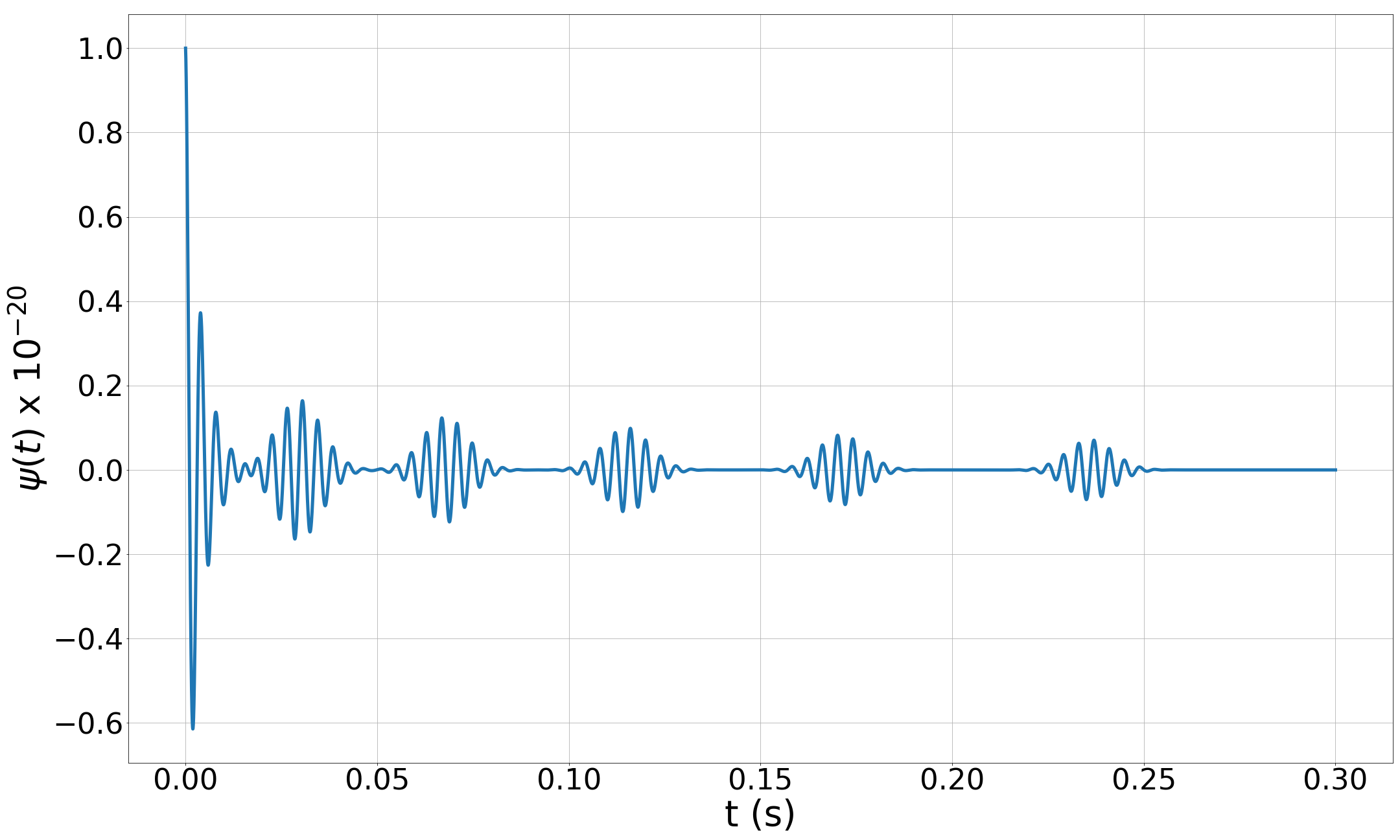}}\end{subfigure}
\caption{GW echoes waveform templates.}
\label{fig1}
\end{figure*}

 \citep{echo1}, in their recent work, have shown that the ringdown part can be used as a probe to study compact objects' structure smaller than that of a BH having a proton-sphere. Such structures will also exhibit similar ringdown features as that of a BH, called Exotic Compact objects (ECO).

Initial ringdown signals, such as those observed by the LIGO detector, reﬂects only the geometry near the photon sphere and is thereby presumed to be very comparable to that of a black hole; if the surface of the ECO is deep inside the photon sphere. GWs, specifically the echoes, reﬂected by the ECO's surface, are expected to show up only in the GW signals at a later stage. Physically, echoes in the late-time ringdown signal are caused by repeated damped reﬂections between the ECO's surface and the potential barrier at the photosphere. The time interval between the two successive echoes governs the system's fundamental physics, and it also leads to sub-classification of the echoes on the ground whether the echoes are at a uniform repeated interval or different interval.

Using the waveform generation approach, as mentioned in \cite{echo5}, we generated a template bank based on the time interval between the echoes that can be equal or unequal. Fig-\ref{fig1} shows examples of the above Constant Interval Echoes (CIE) and Unequal Interval Echoes (UIE) waveform.\footnote{Public repository for producing GW echo templates \url{https://www.darkgra.org/gw-echo-catalogue.html}}

\subsection{\textbf{Core-collapse supernovae} }
\label{2-E}

Core-collapse supernovae (CCSNe) events radiate massive radiation energies in the form of GWs that marks the violent death of massive stars having masses $\sim$ 8M$_\odot$. This collapse will trigger a supernova explosion if the star’s mass is \textless 40M$_\odot$- 50M$_\odot$ (\cite{c1}), resulting in the formation of \texttt{Types II} and \texttt{Ib/Ic} supernovae (\cite{c2}). 

For finding numerical solutions and computation of GWs due to CCSNe, \cite{c4} used quadrupole approximation valid for nearly Newtonian sources. From \cite{58}, $h = \sqrt{ \langle h_{+}^{2}\rangle  +\langle h_{\times}^{2}\rangle}$ , where $\langle \rangle$ indicate that averages taken over wavelength and angle of view at the sky. Corresponding to the $A_{20}^{E2}$ quadrupole amplitude and $\theta$ corresponding the angle of view, GW strain can be calculated using Eq-\ref{eq11} (\cite{c5}). \begin{equation}\label{eq11} h_{+} = \frac{1}{8} \sqrt{\frac{15}{\pi}}\frac{A_{20}^{E2}}{r}\sin ^2{\theta} \end{equation}
We have generated a template bank using pre-simulated data that contains a set of 26 rotational supernova core collapse models\footnote{\url{https://www.mpa-garching.mpg.de/rel_hydro/axi_core_collapse/index.shtml}} in axisymmetry.

\subsection{\textbf{Detector noise}}
\label{2-F}
For ground-based interferometers, it is impossible to insulate the interferometers from the environmental noises completely. The interferometer's ability towards detectable signals is affected by fundamental noise sources such as thermal noise tracing out from the test masses and their suspension systems; quantum noise comprises shot noise at high frequencies and technical noise sources, controlled by the design of the detector. The transient anthropogenic sources, weather, equipment malfunctions, and occasional transient noise of unknown origin also superimpose the GW signals mentioned above. Thus, the signal must undergo a series of signal processing, removal of the unwanted frequencies\footnote{frequency components where noise dominates for example the dominant glitch at 1080Hz (non evolving) $-$ 1080 lines (\cite{g3})}, adding to a computational cost and time before indicating a strong probability of a true GW event (\cite{n1}, \cite{n3}, \cite{n2}).

\subsection{\textbf{Glitches}}
\label{2-G}
Transient non-Gaussian noises that have a significant characteristic, known colloquially as glitches. Glitches are usually technical or hardware noises that trickle into the detector featured by masking the true signal. Gravity Spy\footnote{\url{https://www.zooniverse.org/projects/zooniverse/gravity-spy/about/research}} is a citizen science project aimed to classify the glitches (\cite{g2}). So far, 22 classes of glitches\footnote{while this article was in writing few new glitches have been identified: \url{https://www.zooniverse.org/projects/zooniverse/gravity-spy/talk/762}, although they have not been added to the work} have been classified based on their known features and appearances. Results in \cite{g3} highlight that most of the clusters post-classification are well sparsed, but a few tend to overlap, which is justified by the closeness in the frequency range, power, and appearance.

We have taken $13$ classes of glitches (Table-\ref{table1}) from amongst the 22 classes of glitches as labeled in works of \cite{g3}.\footnote{A few glitches such as due to the raven crossing \big[R. Schofield, \textit{Why the gw channel detects thirsty black ravens along with colliding black holes (2017)}, URL: \url{https://alog.ligo-wa.caltech.edu/aLOG/index.php?callRep=37630.}\big], phone ringing, noise due to aircraft \cite{berger2018identification}, or other such events have a lesser probability of occurrence though have enough power to disturb the signal.}

\begin{table}
\centering
  \caption{Classes of glitches used in our dataset and the specific features under which classification has been done as presented in Bahaadini \textit{et al.} (2018).}
  \label{table1}
\begin{tabular}{cccc}
\hline
Class&Frequency Range(Hz)&Duration&Evolving\\[0.5ex] 
\hline \hline
Low frequency Burst&30-500&Short&Yes\\
Helix&30-500&Short&Yes\\
Koi Fish&30-500&Short&Yes\\
Light Modulations&30-200 (sharp spikes)&Long&Yes\\
Paired Dove&5-150&Short&Yes\\
Power Line&$\sim$60&Short&No\\
Repeating Blips&30-500&Short&No\\
Scattered Light&20-60&Long&Yes\\
Scratchy&30-300&Long&Yes\\
Violin Mode&$\sim$500&Short&No\\
Tomte&30-300&Short&Yes\\
Whistle&$>$300&Short&Yes\\
Wandering Lines&$>$1000&Long&Yes\\
\hline
\end{tabular}

\end{table}
\section{\textbf{Previous works}}
\label{3}
In this section, we discuss the established methods that have been employed for the classification of astrophysical signals. Noise signals usually superimpose astrophysical signals, thus require filtering. Match filtering method is one of the methods wherein the noisy-data post signal processing is compared to pre-simulated templates, which is time-consuming and computationally cumbersome (\cite{pw1}). The introduction of DL in astronomy was primarily employed by Carleo \textit{et al.}. Since then, many DL models have been proposed, primarily to extract astrophysical signals from the noisy data with exceptional accuracy (\cite{g2}, \cite{g3}). The classification of glitches in \cite{g3} is a stark example of how efficiently and accurately the classification can be accomplished. However, a DL model with an arrangement of neurons triggering after some activation threshold makes it much more efficient in feature extraction. CNN method is an acute method for classification problems and has been successfully employed for various other classification domains. The classification done by \cite{pw2} and \cite{pw3} shows how efficiently the signal classification can be done.

Moreover, CNN's have shown high accuracy when predicting the given dataset's multiple features, such as predicting emotion and movie genre prediction. In our work, we have also used a similar approach, a \texttt{multi-label classification} model for predicting the parameter space of the GW event, if detected.

\section{\textbf{Methodology}}
\label{5}
Herein we discuss the data generation scheme for the GW events and the noise segments from the LIGO strain data that have been used for injection and training (as mentioned in Sec-\ref{sec:level2}). We also highlight the \texttt{Image-to-Sound} methodology employed to convert glitches collected in the form of spectrograms to time-series signals. The central reason was to bypass search for glitches over the complete LIGO-run (O1 and O2). Moreover, the Gravity Spy project has a massive dataset for glitches that can be used\footnote{\url{https://zenodo.org/record/1476551\#.YA452OgzZnI}}. We further extend this section towards introducing the DL model employed for prediction of the parameter space, which in our case are 23 labels as listed in Table-\ref{label}. We thereon describe the testing and data validation methodology.

\begin{table}
\caption{List of 23 labels that were used in our dataset. The first six labels are Primary (M$_{1\odot}$) and Secondary (M$_{2\odot}$) masses (in M$_\odot$). Next three labels represent distance ($D_{L}$) in Mpc and the last three correspond to frequencies in Hz (two upper frequency - $f_u$, and one lower frequency ($f_l$, in Hz). For GW Echo class the primary two labels represent UIE and CIE waveforms and the lower two represent the frequencies ($f$, in Hz) at which signals are generated. For CCSNe first two labels after Type-III Supernova represents $r$ in Mpc and last two represents $\theta$. Glitches and Detector noise has been explicitly mentioned.}
\label{label}
\begin{center}
\begin{tabular}{ |c|c| }
Class & Label name (parameter)\\
\hline
\hline
\multirow{12}{*}{BHB / NSB}& 5 (M$_{1\odot}$)\\ & 10 (M$_{1\odot}$)\\& 15 (M$_{1\odot}$)\\& 5 (M$_{2\odot}$)\\& 10 (M$_{2\odot}$)\\& 15 (M$_{2\odot}$)\\& 50 ($D_{L}$)\\& 250 ($D_{L}$)\\& 450 ($D_{L}$)\\& 60 ($f_u$)\\& 120 ($f_u$)\\& 50 ($f_l$)\\ \hline
\multirow{4}{*}{GW Echoes}&UIE (waveform type)\\& CIE (waveform type) \\& 250 ($f$)\\& 280 ($f$)\\ \hline
\multirow{1}{*}{Glitches}&Glitches as mentioned in table-\ref{table1}\\ \hline
\multirow{5}{*}{CCSNe}&III (Supernova type)\\& 10 ($r$)\\& 30 ($r$)\\& 30 ($\theta$)\\& 60 ($\theta$)\\ \hline
\multirow{1}{*}{Noise}&Detector Noise\\
\hline\\
\end{tabular}
\end{center}
\end{table}

\subsection{\textbf{Dataset generation}}
\label{5-A}
Each GW event discussed in Sec-\ref{sec:level2} has been synthetically injected primarily as an individual class to the \texttt{background wave}: time-series data segments sampled out from LIGO strain data possessing features as described in Sec-\ref{5-A-1}. However, there is always a probability of multiple signals being detected in the same sampled time, either clustered or sparsely spread within that sample. Thus, considering the multiple signal probability, the combination of individual GW events (Binaries and CCSNe) have been injected randomly in the sampled background wave. The 12 seconds time sample allowed us to add this \texttt{mixed class} so that transient GW events' signals can be well sparsed within the same sample.

For this work, we have created a dataset with time-series signals of 12 seconds data. Partition has been done for training, testing, and validating into $70\% $, $20\% $, and $10\%$ respectively. Several examples not containing any injections are also a part of the dataset so that the trained model maintains physical reality as far as possible. We generated four times as many examples that contain an injection than a noise sample.

\subsubsection{\textbf{Background wave}}
\label{5-A-1}
Background wave is the LIGO strain data sampled at a particular frequency with no GW detections. We have injected the GW signals into the background wave. A total of $6$ samples were collected such that the following key features are maintained while collecting the data:
\begin{itemize}
\item 	The signal should not have any real detection signal and must only contain detector noise.
\item 	The signal is sampled out must be free from any glitch signal \textit{as far as possible} since some of the glitches are of long duration and non-evolving, for example, 1080 Lines.
\item 	Both the detectors should be functional.
\item   No time segments having hardware injections.
\end{itemize}
The $6$ samples, when injected with the GW events, augmented the number of data samples many folds. Although, the memory restriction forced us to keep the samples as low as $6$. The data samples were collected for GPS time beginning from 1164685312 to 1164685384 in sets of (1164685312, 1164685324), (1164685324, 1164685336), (1164685336, 164685348), (1164685348, 1164685360), (1164685360, 1164685372) and (1164685372, 1164685384).

\subsubsection{\textbf{Blackhole binaries and Neutron star binaries}}
\label{5-A-2}
The basic phenomenology for developing waveform template bank of BH binaries and NS binaries has been discussed in Sec-\ref{2-A}. This set of signals was generated using the PyCBC pipeline. The LALSuite simulations return two polarization modes, namely $h_{\times}$ and $h_{+}$. PyCBC provides routines for calculating these polarized strain values. The table-\ref{table2} summarizes the parameters that we have used to produce a template bank of the desired SNR. The primary (M$_{1}$ in M$_\odot$ referred here as M$_{1\odot}$) and secondary masses (M$_{2}$ in M$_\odot$ referred here as M$_{2\odot}$) ranges from 5 to 15 with intervals of 5M$_\odot$. Luminosity distances are at intervals of 200 Mpc (referred as $D_{L}$), beginning from 50Mpc to 450 Mpc, with upper frequency ($f_u$) varying between 60Hz and 120Hz with a fixed upper frequency ($f_l$) at 50Hz resulting in a total of $54$ signals from individual approximant. Using $5$ approximants, we generated a total dataset of $270$ imposed on $6$ background waves resulting in $1620$ binary signals. The parameters have been arranged in Table-\ref{label}.

\begin{table}
\centering
\caption{The parameters used to prepare a template bank for BHB and NSB. The waveforms have been generated using methods provided in PyCBC. All the waveforms generated do not posses spin parameter ($S_x = S_y = S_z = 0$).}\label{table2}
\begin{tabular}{cccccc}
\hline
Approximant&M$_{1\odot}$&M$_{2\odot}$&$D_L$ (Mpc)&$f_u$(Hz) &$f_l$(Hz)\\
\hline   \hline
TaylorT1&5,10,15&5,10,15& 50-450&60,120&50\\ 
TaylorT2&5,10,15&5,10,15& 50-450&60,120&50\\
EOBNRv2&5,10,15&5,10,15&  50-450&60,120&50\\
SEOBNRv1&5,10,15&5,10,15& 50-450&60,120&50\\
SEOBNRv2&5,10,15&5,10,15& 50-450&60,120&50\\
\hline
\end{tabular}
\end{table}

\subsubsection{\textbf{GW echoes}}
\label{5-A-3}
GW echoes were simulated using mathematical routines, as discussed in Sec-\ref{2-D} (\cite{echo5}). The model parameters that were altered to produce a template bank are described in this section. The waveforms that have been generated do not contain a GW transient before the echo signal, unlike the physical scenario wherein the GW echo follows a GW transient. The idea was to allow the DL model to be able to differentiate between the parameters of the echo signal. Waveform templates have been generated referring to the symbols used in \cite{echo5}. Value of $\tilde N_{echo}$ was varied from 0 to 8 for two sets: UIE and CIE. Keeping the value of $r$ fixed at $0.3$, the dataset was generated for two frequencies: $250$Hz and $280$Hz. The simulation was done for $10$ iterations\footnote{if we choose a single segment of “background wave" and a single GW echo signal for a given set of parameters injection was done on 10 copies of the same “background wave” with same GW echo signal at different positions of the signal}, randomly injecting echo signals to the $6$ background wave samples resulting in the total number of $1920$ samples.

\subsubsection{\textbf{Core Collapse Supernova Explosion}}
\label{5-A-4}

We have generated a template bank using pre-simulated data containing 26 rotational core-collapse supernova models\footnote{\url{https://wwwmpa.mpa-garching.mpg.de/rel_hydro/axi_core_collapse/}}. The data obtained encapsulated time-series evolution of $A_{20}^{E2}$. The data was converted in terms of two GW polarization ($h_+, h_\times$), time-series signals using the Eq-\ref{eq11}, the angle of view $\theta$ varying at 30$^{0}$ and 60$^{0}$ and distance, $r$ varying from 10 Mpc and 30 Mpc for a Type III supernova resulting in a total of $624$ signals when injected to the background wave. An example of the waveform has been shown in Fig-\ref{fig9}.

\subsubsection{\textbf{Glitches} }
\label{5-A-5}

\begin{figure}
\centering
\includegraphics[width=8cm,height=6cm]{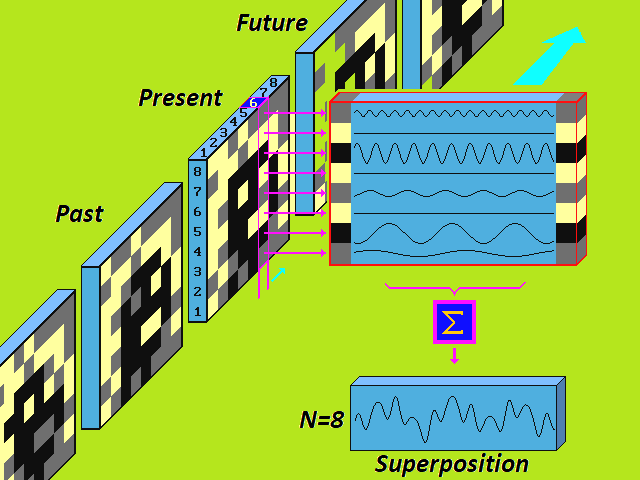}
\caption{In order to achieve a high resolution, each image is transformed into a time-multiplexed auditory representation. Each column's pixel excites a similar sinusoidal wave. This mapping translates for individual pixel, vertical position into frequency, horizontal position into time-after-click, and brightness into amplitude, which is the power in any spectrogram. Here N$=8$ represents the 8$\times$8 dimensions of the image.}
\label{fig3}
\end{figure}

To generate glitches as a time-series sequence, we used a methodology of \texttt{Image to sound mapping}. We employed this method to bypass the search over the complete run of the LIGO strain data for suitable glitch candidates. From the Gravity spy project, we obtained the glitch signals in the form of spectrograms, and using \texttt{Image to sound} mapping; we successfully converted spectrogram to time-series signals. Since the input data (image set) is already whitened, the absolute amplitude cannot be inferred directly and thereby require normalization. The normalized signals are thereon rescaled to be at a detection threshold of SNR>8. This value is arbitrarily chosen and has previously been used in \cite{AAAA} and \cite{BBBB}. We generated short signals, as mentioned in table-\ref{table1} from a methodology of Image to sound mapping, wherein every image having color components as RGB or grayscale can be transformed into respective sound. The short-duration glitches are harder to detect in the framed sample, and few are shorter than a second. Fig-\ref{fig3} illustrates the conversion procedure of a three grey-tone image of dimensions $8\times 8$. This mapping translates for individual pixel, vertical position into frequency, horizontal position into time-after-click, and brightness into amplitude, which is the power in any spectrogram. Each column's pixel is used to excite a similar sinusoidal wave in the audible frequency range.
 
The simplest form results in a spectrogram synthesis where every pixel is mapped to a brief pure tone of a specific frequency and amplitude corresponding to that pixel's power (brightness). However, a pure tone of less than infinite duration (theoretically) does not give a perfect spectral line but evaluates a spectral envelope with the main lobe and side lobes. Thereon the spectral envelope will overlap, known as spectral leakage. Thus, the shorter duration sounds will have a greater spread of frequency, limiting the sufficient resolution in spectrogram-like mappings, resulting in a signal that is not usually the \texttt{pure sound signal} having some random noise that needs processing before injecting into the background wave. To filter the signal and to obtain the \texttt{true} glitch signal, we used \texttt{Butterworth} and a \texttt{Median} filter of Scipy Python package\footnote{\url{https://docs.scipy.org/doc/scipy/reference/generated/scipy.signal.filtfilt.html}} \texttt{scipy.signal.filtfilt} \citep{2020SciPy-NMeth}. Thus time-series signal of the glitches as mentioned in table-\ref{table1}:  helix, koi fish, light modulation, low-frequency burst, paired dove, repeating blip, scattered light, tomte, scratchy, whistle, wandering line, and violin-mode were generated using this method. In Fig-\ref{fig8} we have shown the time-series glitch signals obtained using this method.

\subsubsection{\textbf{Mixed Dataset}}
\label{5-A-6}
This part of the dataset has been introduced purposely to highlight the non-zero probability of multiple signals present in the time sampled. The random injection ensures that the signal does not occupy the central vector space and is sparsed. There are two sets of mixed signals: \texttt{Mix of binaries}- having random injections of  BHB, NSB, GW echoes, \texttt{Mix of CCSNe}- CCSNe and Glitches.

\subsection{\textbf{Dataset Preparation}}
\label{5-AB}
\begin{figure}
\centering
\subfigure{\includegraphics[width=4cm, height=5cm]{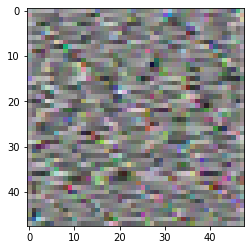}}
\subfigure{\includegraphics[width=4cm, height=5cm]{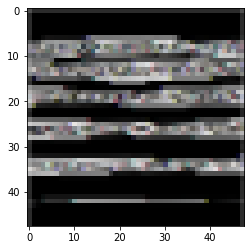}}
\caption{Sample image of shape $48\times48\times3$, a BH binary and GW echoes, used for training. Initially a 1D data it has been reshaped as $48\times48\times3$ 2D image or in terms of DL a $3$ dimensional tensor.}
\label{new}
\end{figure}

The 1D time-series data was reshaped to a 48$\times$48$\times$3 shape. Fig-\ref{new} represents prepared image. Labels were generated as mentioned in Table-\ref{label} along with the generation and injection of the signals. Labels prepared are one-hot-coded, which means that each label, as mentioned in Table-\ref{label} has either a 0 or 1 value; 1 corresponding to the presence of the parameter and 0 the absence. For, e.g., in case of BHB using TaylorT1 approximant with M$_{1\odot}$ as 5 and M$_{2\odot}$ as 15 with $D_L$ as 250 Mpc having $f_u$ 120 and $f_l$ 50 and Noise (background wave) the labels corresponding to these parameters were tagged 1 and others 0 marking the absence of those parameters. For signals that had no injection, only the Noise label was tagged 1, others as 0\footnote{If the injections are made on the “background signal” that is not labeled as “Noise” since that signals segment \big[background signal + injected strain\big] contains some GW information as well. The same applies to the Glitch class as well; \big[background signal + injected strain of glitch\big] has been labeled as "Glitch" and not as "Noise."}.

\subsection{\textbf{Model Selection}}
\label{5-B}
Due to the nature of the dataset, CNN has been employed to extract 12 features of time-series data. Compared to its predecessors, CNN's ability to automatically detect the essential features without any human supervision makes it the apt model for our dataset. DL models have a structured arrangement of layers with several neurons that trigger only after some activation threshold, making it much more efficient in feature extraction and controlling which feature needs to be extracted. CNN-based models have been previously used for several tasks: Natural Language Processing (NLP), Image classification, Glitch classification, GWs searches as well (\cite{g2}, \cite{g3}, \cite{pw3}). As mentioned earlier, CNN's have performed extremely well on binary classifications; however, earlier from \cite{ms2} and works in biomedical areas \cite{ms1} indicates the ability to perform on a similar level for a multi-classification problem.

\subsection{\textbf{Model Architecture}}
\label{5-C}
\begin{figure*}
\centering
\includegraphics[width=18cm, height=7.5cm]
{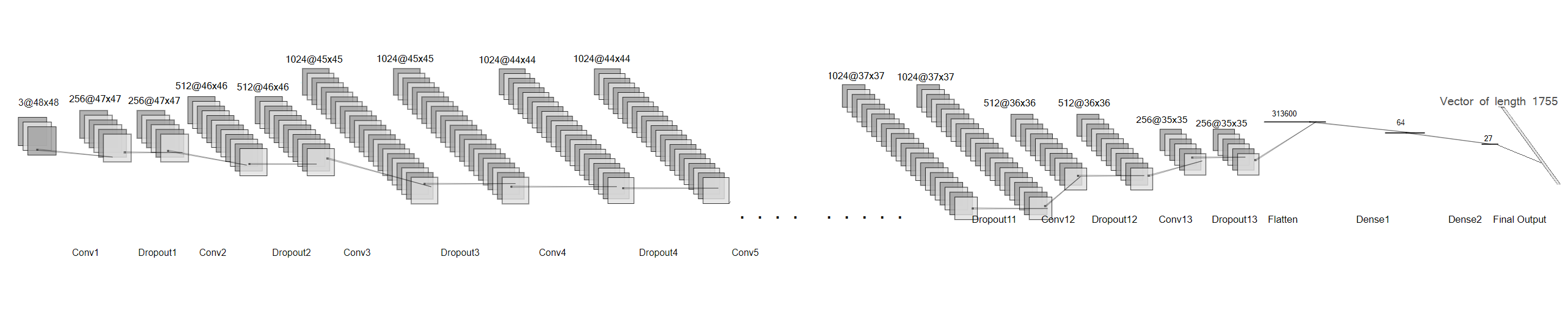}
\caption{The sequential CNN model architecture with 13 CNL, 13 Dropout layers, 1 Flatten layer, and 2 Dense layers. All the layers and filters have not been shown and are is described in Table-\ref{table5}}\label{model}
\label{fig5}
\end{figure*}

From Fig-\ref{fig5}, it can be seen that our model follows a sequential approach with tensors passing through various layers sequentially. The network is 29 layers deep with 13 Convolutional layers (CNL), 13 Dropout layers, 1 Flatten layer, and 2 Dense layers. We defined a nonlinear activation function after each CNL, ReLU, followed by a dropout layer. We avoided Max-pooling or Average-pooling primarily to prevent loss of information since these layers tend to reduce the number of training parameters and the computational load. With 30\% of neurons activating after each layer, we used 2 Dense Layers with ReLU and Sigmoid activation, respectively. We have conducted the training experiment ten times corresponding to each varying parameter listed below. The results obtained on altering various parameters can be summarized from Fig-\ref{fig6}.

\begin{itemize}
\item \texttt{learning rate}: $10^{-1}$,$10^{-2}$, $10^{-3}$
\item \texttt{batch size}: 100, 200, 250, 500
\item  \texttt{Kernel Size}: 2, 3, 4
\end{itemize}
Finally from the summary of Fig-\ref{fig6} and Fig-\ref{fig7} we chose our network architecture as mentioned in table-\ref{table5}.

During the first convolution, the input signals converted to the shape (48, 48, 3) are primarily convoluted by the 256 filters having kernel size as (3, 3) and translating 1 step at a time. The architecture grows and widens to 1024 layers from 256 to 512 and then to 1024, and then narrows down to 256 filters at the last CNL. A Flatten layer follows this, and finally, we have two dense layers giving the outputs as an array of (64) values and then as (23) respectively. The output 1D network output is then passed through a sigmoid layer, mapping it within the interval (0,1).  These 23 values are the predicted parameter space of the input signal. This method yields a relatively large receptive field for 12 seconds data with a depth of 13 CNL blocks. However, memory limitations during training upper-bounded the number of channels to 1024. This implementation has been done in (in Python-3.6.7) is based on the Keras\footnote{\url{http://keras.io/}} deep learning framework (\cite{keras}). A schematic model architecture is depicted in Fig-\ref{model}.

\begin{table*}
\centering
\caption{The sequential CNN model architecture with 13 CNL, 13 Dropout layers, 1 Flatten layer, and 2 Dense layers. All the layers and filters have not been shown. Here we also highlight the input and output shapes after each layer. The flow of data and how probability for each of 23 labels are derived can be drawn from this table.}\label{table5}
\begin{tabular}{ccccccc}
         Layer&Input Shape&Output Shape&kernel size&stride&padding& Activation\\
         \hline \hline\\
        Convolution$^{1}$&(48, 48, 3)&(47, 47, 256)&(3, 3)&1&2&ReLU\\
        Dropout$^{1}$&(47, 47, 256)&(47, 47, 256)&-&-&-&-\\
        Convolution$^{2}$&(47, 47, 256)&(46, 46, 512)&(3, 3)&1&2&ReLU\\
        Dropout$^{2}$&(46, 46, 512)&(46, 46, 512)&-&-&-&-\\
        Convolution$^{3}$&(46, 46, 512)&(45, 45, 1024)&(3, 3)&1&2&ReLU\\
        Dropout$^{3}$&(45, 45, 1024)&(45, 45, 1024)&-&-&-&-\\
        Convolution$^{4}$&(45, 45, 1024)&(44, 44, 1024)&(3, 3)&1&2&ReLU\\
        Dropout$^{4}$&(44, 44, 1024)&(44, 44, 1024)&-&-&-&-\\
        Convolution$^{5}$&(44, 44, 1024)&(43, 43, 1024)&(3, 3)&1&2&ReLU\\
        ...&...&...&...&...&...&...\\
        ...&...&...&...&...&...&...\\
        ...&...&...&...&...&...&...\\
        Dropout$^{11}$&(37, 37, 1024)&(37, 37, 1024&-&-&-&-\\
        Convolution$^{12}$&(37, 37, 1024)&(36, 36, 512)&(3, 3)&1&2&ReLU\\
        Dropout$^{12}$&(36, 36, 512)&(36, 36, 512)&-&-&-&-\\
        Convolution$^{13}$&(36, 36, 512)&(35, 35, 256)&(3, 3)&1&2&ReLU\\
        Dropout$^{13}$&(35, 35, 256)&(35, 35, 256)&-&-&-&-\\
        Flatten&(35, 35, 256)&(313600, 1, 1)&-&-&-&-\\
        Dense$^{1}$&(313600, 1, 1)&(64, 1, 1)&-&-&-&ReLU\\
        Dense$^{2}$&(64 ,1, 1)&(23, 1, 1)&-&-&-&Sigmoid\\
\hline
\end{tabular}
\end{table*}

\begin{figure*}
\centering
\includegraphics[width=18cm, height=8cm]
{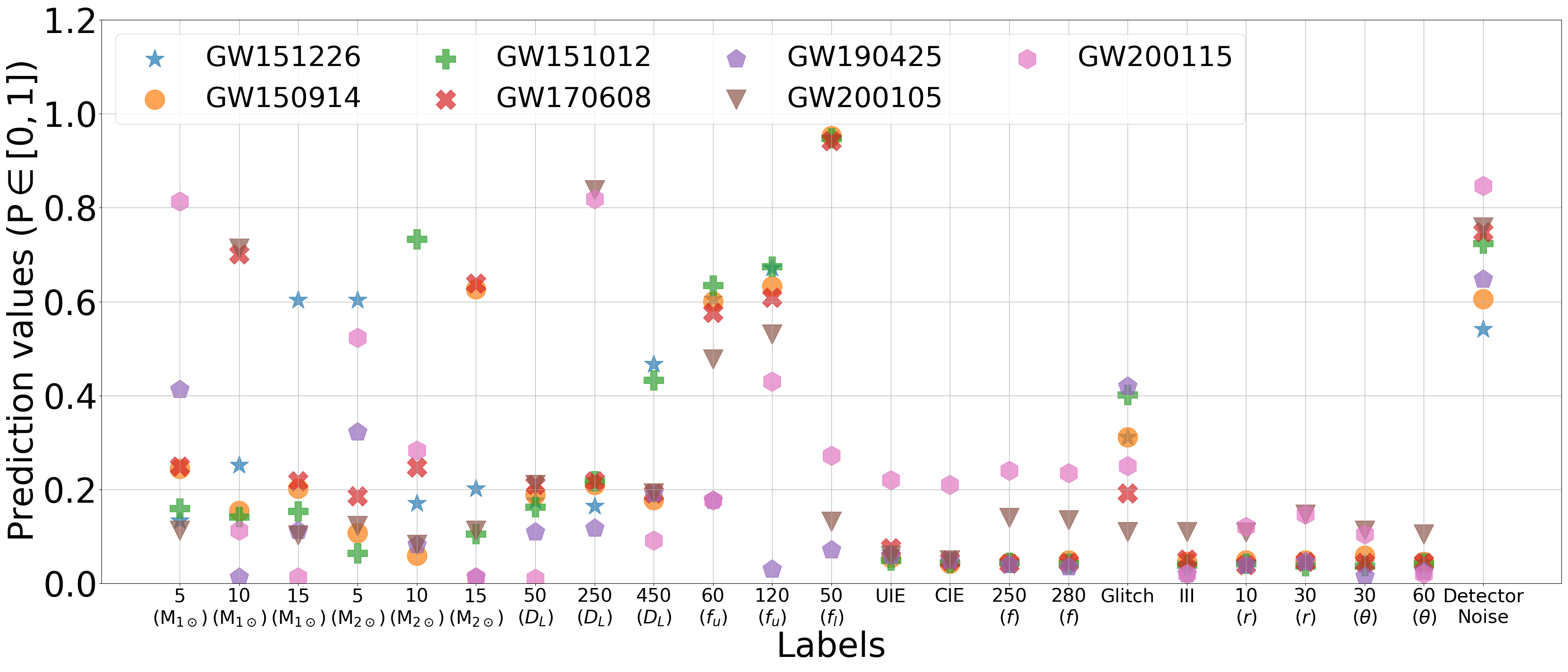}
\caption{Prediction values for the already detected GW events. On horizontal axis the labels are marked from Table-\ref{label} and correspondingly the normalized prediction values are plotted against each label. The values within "( )" represent the parameters plotted.}\label{valid}
\end{figure*} 

\begin{figure*}
\centering
\includegraphics[width=18cm, height=8cm]
{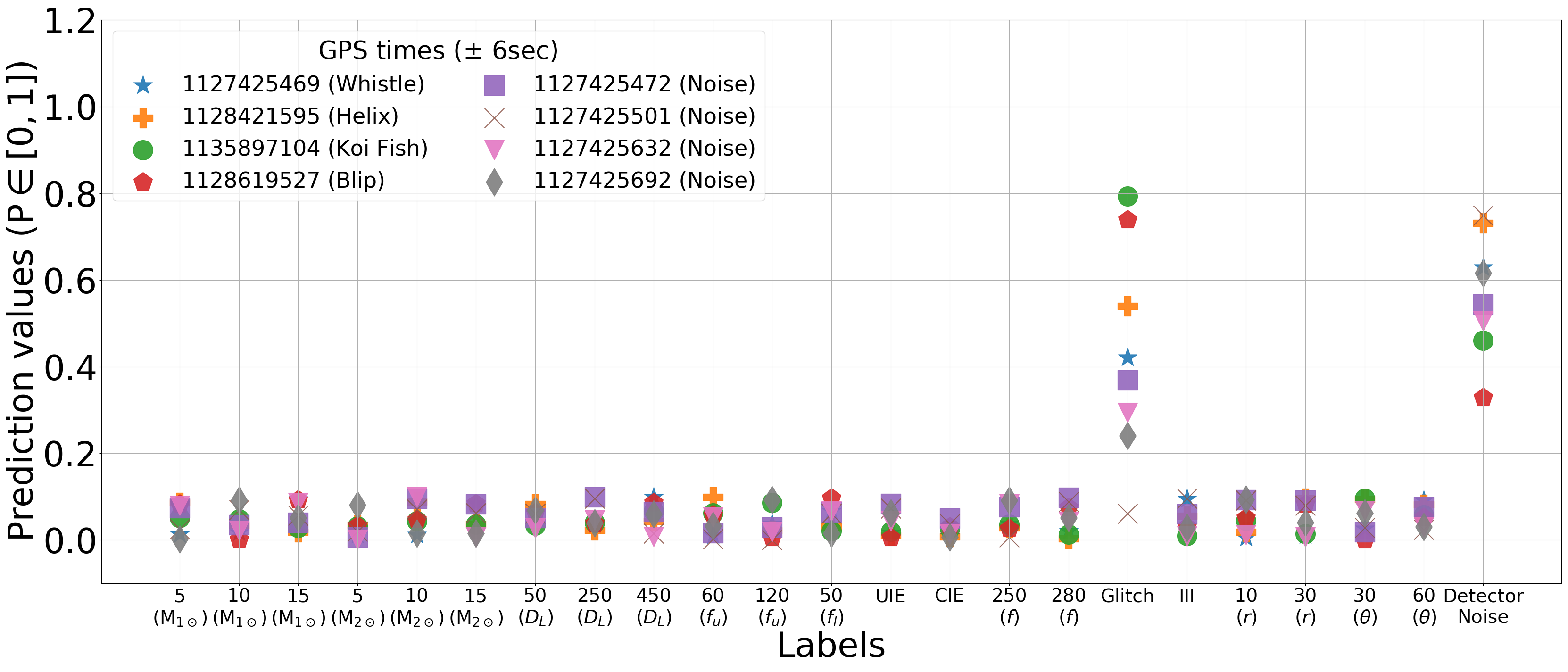}
\caption{Prediction values for the glitch class and noise class. Four glitch classes: Whistle, Helix, Koi Fish, and Blip, have been shown. The event's time has been highlighted, and the $\pm 6$sec of time-series data around the highlighted time has been used. Similarly, the central time has been highlighted for noise signal, and $\pm6$sec around the central time has been used for validation.The values within "( )" represent the parameters plotted.}\label{valid_noise}
\end{figure*} 

\subsection{\textbf{Model Training}}
\label{5-D}

Curriculum learning has been employed to train the model. Initially, the DL model was trained over easier datasets that had a single signal or no injection. Finally the training was concluded on \texttt{mixed-dataset} (\cite{cirr1}, \cite{cirr2}). For our dataset during training, the performance was regulated by continually estimating the validation set model. During training, the kernel entries are optimized by Stochastic Gradient descent (with momentum 0.95), having a learning rate of $10^{-3}$ and decay $10^{-6}$. With every epoch, the dataset is and divided into batches of batch size 500. Binary Cross-Entropy (BCE) has been used, calculated batch-wise at every time step. This calculated batch loss gradient is evaluated for all kernels, and error back-propagation is used to tweak the kernel values towards a converging solution. At the end of every epoch, the network's performance with its current parameter values is evaluated on the full training and validation data set. The model's training was done on a Portable machine with Intel Core-i7 7700 HQ processor (base clock 2.8 GHz and boost clock 3.6 GHz), NVIDIA GTX 1060 6 GB graphics, with 16 GB RAM, and took approximately 40 hours to train for 30 epochs. The model's training accuracy marginally increased on further training, but the validation accuracy and testing accuracy decreased, indicating over-fitting.

\section{\textbf{Performance Metrics}}
\label{6}

\subsection{\textbf{Results and discussion}}
\label{6-B}After 30+ hours of training, the model training accuracy, i.e., before the over-fitting, was obtained between 89.97\% and 90.1\%, experimented ten times with the same model architecture. The validation accuracy was $\sim 89.03\%$, and validation loss was  $\sim0.3\%$. We tested the model accuracy on the $20\%$ previously partitioned dataset, resulting in an accuracy of $89.93\%$. Fig-\ref{fig6} represents the parameters we altered during our experimentation and the maximum, which was achieved by the current architecture.

While experimenting with various model parameters, we observed an increase in training accuracy for odd-numbered kernel filter-size along with increasing batch-size. Since we experimented with four batch sizes: 100, 200, 250, 500 sparsely separated, the result cannot be generalized. Also, the batch size from which the inflection in accuracy begins is not acknowledged.

\subsection{\textbf{Result validation}}
We have validated our model using four GW detections which are: GW150914 (\cite{det2}), GW151226 (\cite{det1}), GW170608 (\cite{det3}), GW151012 (\cite{det4}), and GW190425 (\cite{Abbott_2020}), GW200105 (\cite{Abbott_2021}), and GW200115 (\cite{Abbott_2021}). Fig-\ref{valid} shows the predicted values by the proposed DL model corresponding to the 23 labels that are listed in Table-\ref{label}. The parameter space of our label set is very sparse and limited. For BHB and NSB, the lower mass limit is 5M$_{\odot}$ whereas the upper mass limit is 15M$_{\odot}$ with $D_L$ ranging from 50Mpc to 450Mpc varying coarsely\footnote{The mass parameters are evaluated in the detector frame; to convert to the source frame divide by $(1 + z)$, where $z$ is the source redshift}. Also, we have not accounted for the spin, eccentricity, and various other extrinsic parameters that govern the dynamics of the binaries and thus the generated waveform. From the experiments we conducted and given the input datasets the DL model was trained on, we concluded a probability of $\gtrsim60\%$ indicates a strong prediction corresponding to a given parameter. Although this strong prediction does not necessarily indicate correct or incorrect prediction, which solely depends on how well the model has been trained and how close the predicted parameter is to the injected values; this is regarding how confidently the DL model is predicting that parameter.

Referring to the parameters of GW151226\footnote{\url{https://www.gw-openscience.org/events/GW151226/}} that matches our labels: M$_{1}$ as $15.5^{+8.3}_{-3.7}$M$_\odot$ and  M$_{2}$ as $8.2^{+2.3}_{-2.3}$M$_\odot$, with $D_L$ of about $440^{+180}_{-190}$Mpc the model predicts M$_{1\odot}$ and M$_{2\odot}$ with an probability $> 60\%$ for our label of 15M$_{1\odot}$ and 5M$_{2\odot}$ respectively. The sparsity of the label set has restricted a very high probability. Moreover, the DL model predicts the $D_L$ of 450Mpc with a probability $> 60\%$. The frequency increased from 35Hz to 450Hz, for which the model generated slightly inaccurate predictions. This confusion by the model is highlighted with all the four validating GW events: $f_l$, which is fixed at 50Hz $\sim 90\%$ and $f_u$, which are 60Hz and 120Hz to be $\sim 60\%$. For other parameters M$_{1}$, M$_{2}$ and $D_L$ from our labels, the prediction probability is as low as $20\%$, marking the absence of that parameter. Also, it can be noted from Fig-\ref{valid} that parameters of GW echoes and CCSNe have been predicted with $< 5\%$ probability in the signal showing strong confidence in the model towards the ability to differentiate between different classes of GW events. We also observe a spike for the Glitch class but is restricted under $40\%$ for all validating sets. Since signals injected on the background wave is dominated by noise, the prediction of the Noise label is close to $60\%$

For GW150914\footnote{\url{https://www.gw-openscience.org/events/GW150914/}} M$_{1}$ as $38.9^{+5.6 \pm 0.6}_{-4.3 \pm 0.4}$M$_\odot$ and M$_{2}$ as $31.6^{4.2\pm0.1}_{-4.7\pm0.9}$M$_\odot$, with $D_L$ of about $420^{+150}_{-180}$Mpc. The model predicts M$_{1\odot}$ with a probability $< 25\%$ and M$_2 >60\%$, which is a loose prediction since both masses lie outside the domain of $5$M$_{1\odot}$ and $15$M$_{2\odot}$ from our dataset also the $D_L$ is not predicted accurately. Thus a sparsed label set highlights the failure modes of our DL model. The same loose prediction extends in predicting the parameters: $f_l$ and $f_u$. However, the model efficiently understands the distinction between various GW classes and has a $< 5\%$ prediction probability. The trend for glitch and noise prediction is maintained.

In case of GW170608\footnote{\url{https://www.gw-openscience.org/events/GW170608/}} M$_{1}$ as $12.8^{+7}_{-2}$M$_\odot$, M$_{2}$ is $7.5^{+2}_{-2}$M$_\odot$ with $D_L$ of about $320^{+120}_{-110}$Mpc M$_{1\odot}$ and M$_{2\odot}$ with an probability $> 60\%$ for our label of 10M$_{1\odot}$ and 15M$_{2\odot}$. Moreover, the DL model is unable to predict the $D_L$ accurately will all the predictions $< 25\%$. The frequency increased from 35Hz to 450Hz, for which the model generated slightly inaccurate predictions. With $f_l$, which is fixed at 50Hz $>90\%$ and $f_u$, which are 60Hz and 120Hz to be $>55\%$ and $60\%$ respectively. For the prediction probability is as low as $10\%$ marking the absence of that parameter.

For GW151012\footnote{\url{https://www.gw-openscience.org/GWTC-1/}} M$_{1}$ as $28.2^{+14.0}_{-5.5}$M$_\odot$, M$_{2}$ as $16.5^{+4.1}_{-4.8}$M$_\odot$ with $D_L$ of about $1060^{+540}_{-480}$Mpc the model predicts M$_{1\odot}$ with an probability $< 20\%$ and M$_2 \sim70\%$ for a label 15M$_{2\odot}$ of which is a close prediction for M$_{2\odot}$. Since M$_{1\odot}$ and $D_L$ lies outside the domain of study the prediction accuracy is low.

Other GW events:  GW190425\footnote{\url{https://www.gw-openscience.org/eventapi/html/O3_Discovery_Papers/GW190425/v1/}}, GW200105\footnote{\url{https://www.gw-openscience.org/eventapi/html/O3_Discovery_Papers/GW200105_162426/v1/}} and GW200115\footnote{\url{https://www.gw-openscience.org/eventapi/html/O3_Discovery_Papers/GW200115_042309/v1/}} have also been added for the validation wherein we can see the similar results. The GW parameters close to the trained parameters have been predicted with a strong accuracy alongside being able to differentiate between various classes of data. The ability to differentiate between various classes (as mentioned in Table-\ref{label} can be concluded from Fig-\ref{valid} and Fig-\ref{valid_noise}. Referring to Fig-\ref{valid}, one can observe that it does not predict any class from amongst GW echoes or CCSNe. Although we have not tested our model corresponding to CCSNe or GW echoes since we have not yet found any plausible GW candidate to study. Although of extreme interest, CCSNe events are pretty rare, with approximately 2-3 occurring per century in our Galaxy (\cite{CCSNe_detect}). Also, a search attempt by \cite{wang2020searching} for GW Echoes showed no statistically significant evidence. Similarly, in Fig-\ref{valid_noise} the GW parameters have not been predicted with strong accuracy. A clear observation can be drawn from Fig-\ref{valid_noise} wherein the predicted accuracy has been plotted for noise and glitch class. In the cases wherein the probability for the glitch is high, the noise probability is low (but prominent), whereas vice-versa for noise class highlighting the model's ability to differentiate between prominent noise sources (glitches) and random noise. These kinds of multi-label classifications are difficult to get high accuracies for all the labels together, often, some labels are predicted with higher accuracy, and others lag.

\section{\textbf{Conclusion}}
\label{7}
In this work, we introduce the methodology of synthetic preparation of data using NR for BH, NS binaries, and CCSNe. We further investigated the GW Echos and were able to synthetically inject the proposed two categories of the signal as in \cite{echo3}. We carried out an in-depth analysis of deep CNNs for multi-label prediction to predict the parameters of the strain data. Firstly, we critically examined the existing classification methods opted and well tested, thereon attempting to extend classification problems towards multi-label prediction.

We use a simple but subtle network architecture tailored towards the multi-label prediction for the realtime parameter prediction of the strain data. The model does face accuracy loss due to scattered labels, a small dataset required for extraction of 12 features, slightly unbalanced data, and memory upper bound. Thereby, we showcase a selection of \texttt{failure modes} of our model, typical for deep convolutional neural networks.

While preparing the dataset for training, we infer that all injected inputs are physically possible. Future highly sensitive GW detectors could investigate the BBH, BNS, CCSNe, and GW echoes and stochastic sources simultaneously, giving us an ameliorated picture of astrophysical phenomena in the observable universe. With the ever-escalating template space of matched filtering, it would become increasingly challenging to denoise the future GW signals without exponential computing performance. Deep learning models like ours could identify a plethora of different types of such signals much faster than Matched filtering, which would, in turn, allow for faster parameter estimation and aid in the localization of GWs for Multi-Messenger Astronomy. CNNs are an assuring tool for GW astronomy; however, their exact interpretation requires great care and attention, and one model which may work for a specific problem may not work for other challenges.

\section*{Acknowledgements}
This research has made use of data, software, and/or web tools obtained from the Gravitational Wave Open Science Center (\url{https://www.gw-openscience.org}), a service of LIGO Laboratory, the LIGO Scientific Collaboration, and the Virgo Collaboration. LIGO is funded by the U.S. National Science Foundation. Virgo is funded by the French Centre National de Recherche Scientifique (CNRS), the Italian Istituto Nazionale della Fisica Nucleare (INFN) and the Dutch Nikhef, with contributions by Polish and Hungarian institutes. We wish to show our gratitude to many people without whom this work would not have been complete. We also wish to express our gratitude towards all the reviews that were crucial for this work.

\section*{Data Availability}
The code sample and pipelines we used during our data generation scheme are available on PyCBC - Free and open software to study gravitational waves\footnote{\url{https://pycbc.org/}}. The code(s) and the algorithm that has been employed in generating the dataset are available on GitHub\footnote{\url{https://github.com/SSingh087/DL-estimating-parameters-of-GWs}}.

\bibliographystyle{mnras}
\bibliography{main}

\begin{figure*}
\centering
\subfigure{\includegraphics[width=8.5cm, height=8cm]{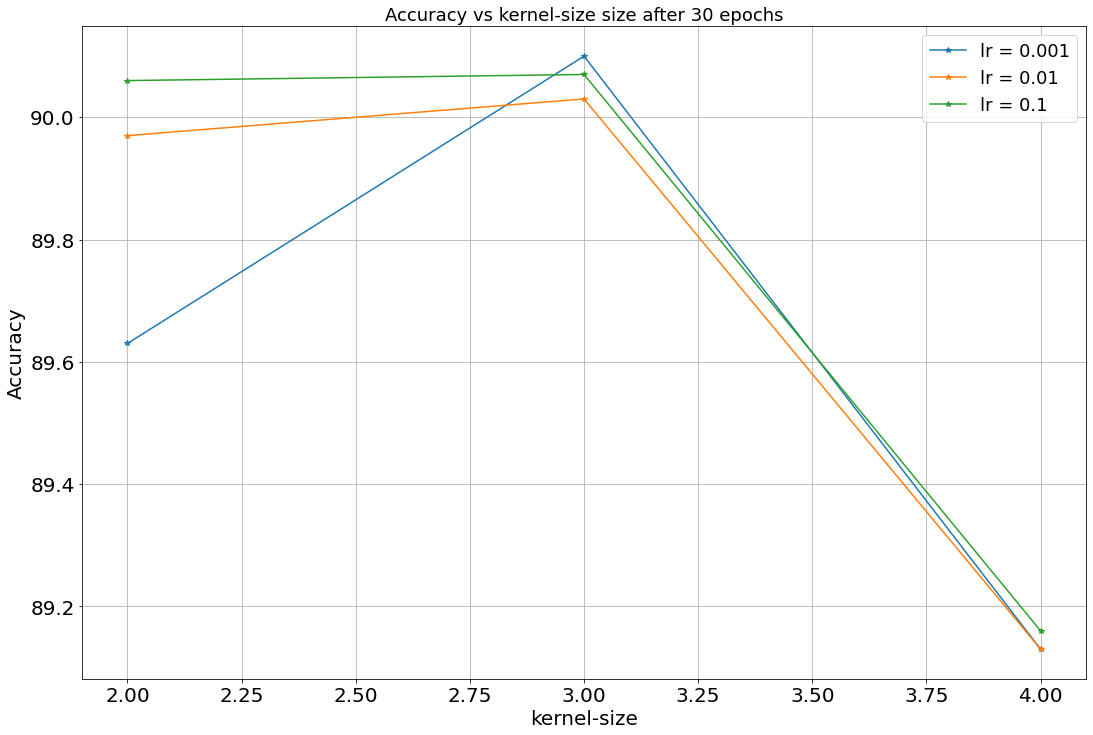}}
\subfigure{\includegraphics[width=8.5cm, height=8cm]{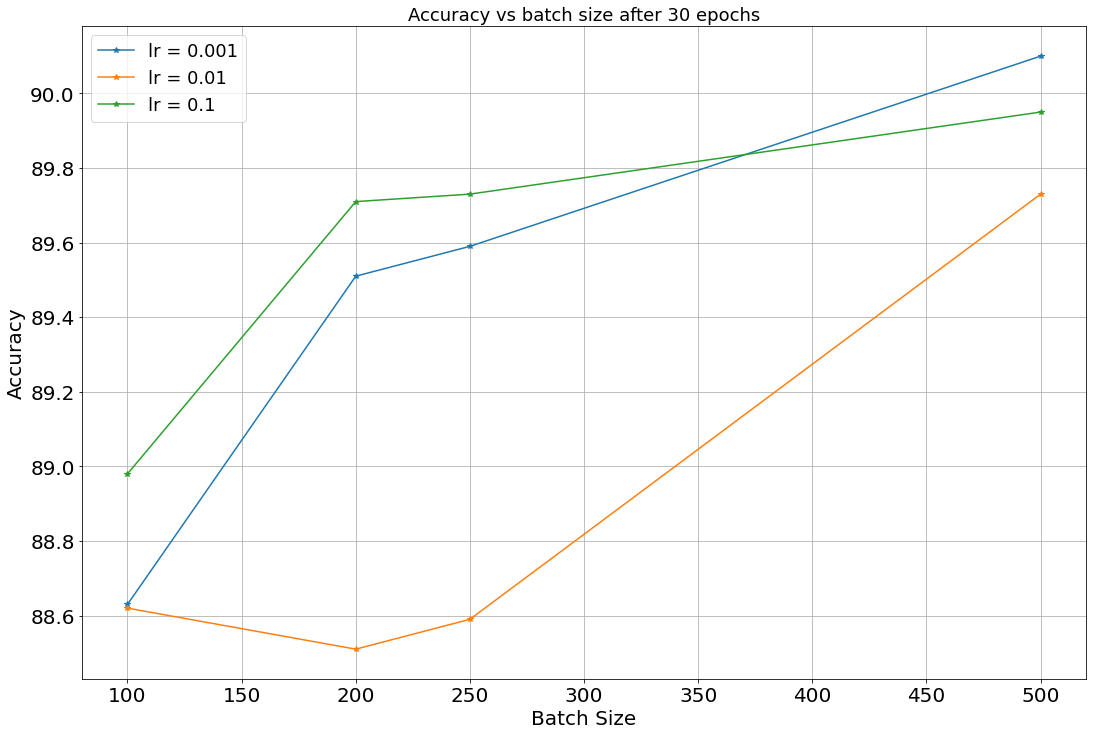}}
\caption{Performance of accuracy with changes in the model parameters during training. The left column represent the accuracy variation with variation in kernel-size, the right column represents accuracy variation with variation in batch-size. The experiments were conducted for three learning-rates: $10^{-1}$, $10^{-2}$, $10^{-3}$.}
\label{fig6}
\end{figure*}

\begin{figure*}
\centering
\subfigure{\includegraphics[width=8.5cm, height=8cm]{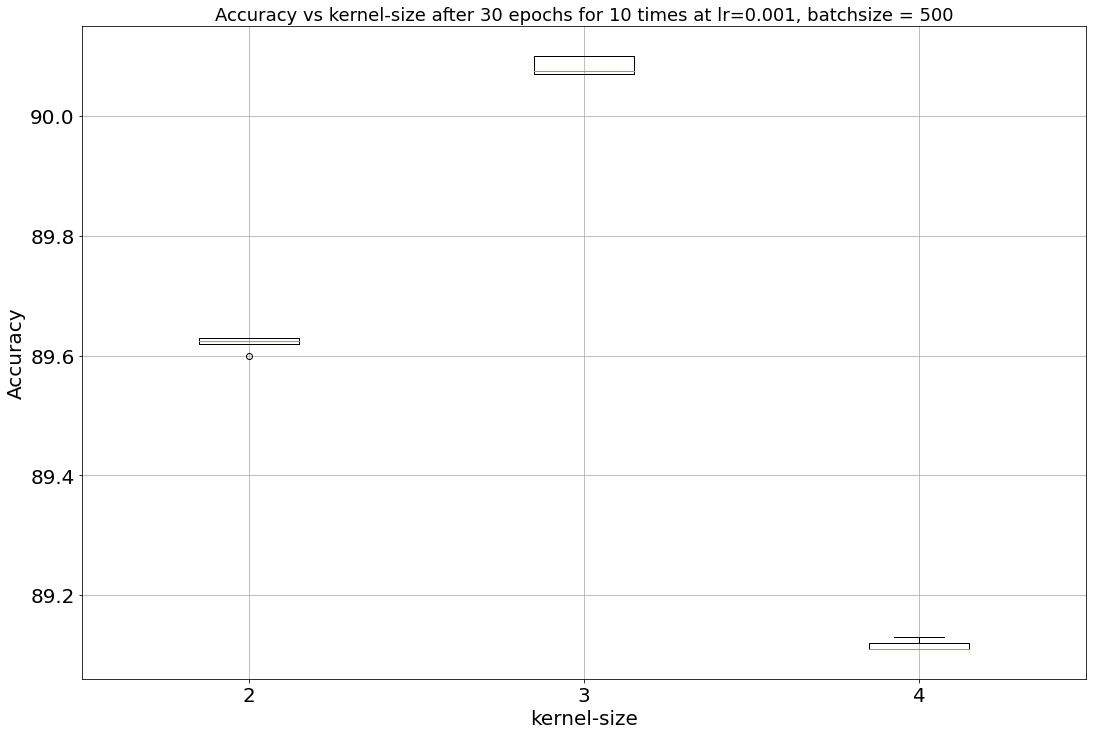}}
\subfigure{\includegraphics[width=8.5cm, height=8cm]{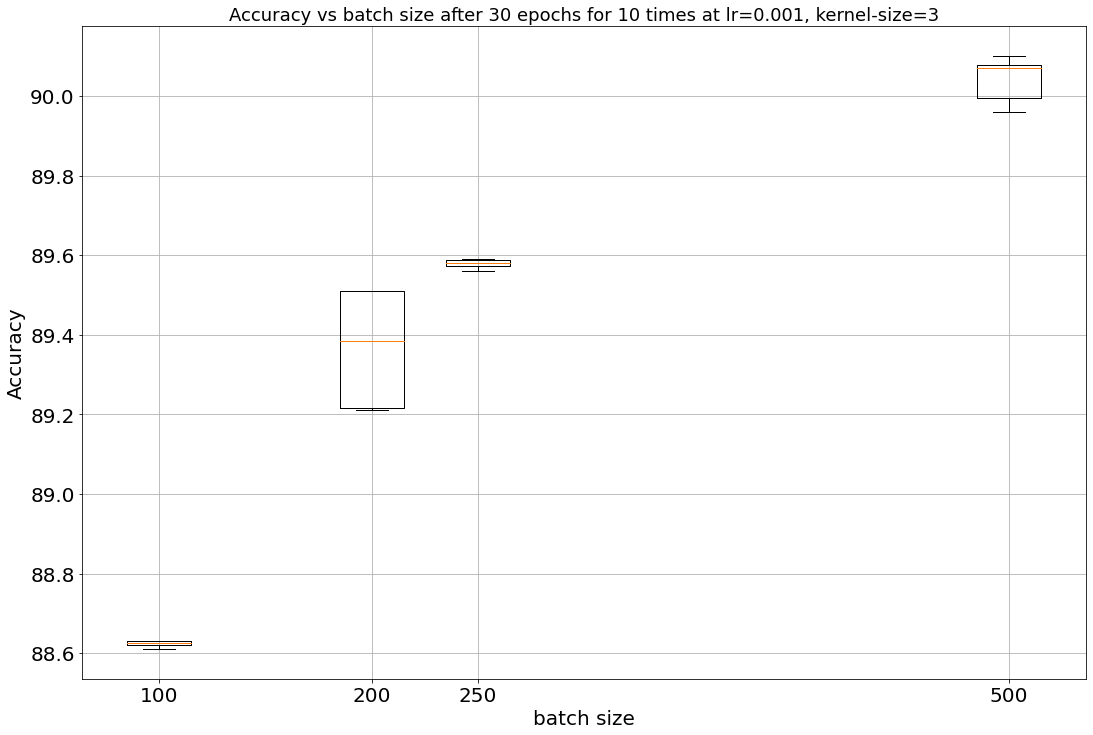}}
\caption{Performance of accuracy with changes in the model parameters during training for 30 epoch ran 10 times. The left and right column represent the same parameters as in Fig-\ref{fig6} experimented for 10 times. Each boxplot represents the spread in the accuracy corresponding to variation of model-parameters}\label{fig7}
\end{figure*}

\begin{figure*}
\centering
\setkeys{Gin}{width=2in,height=2cm}
\begin{subfigure}[]{}
            \includegraphics[width=6cm,height=4cm]{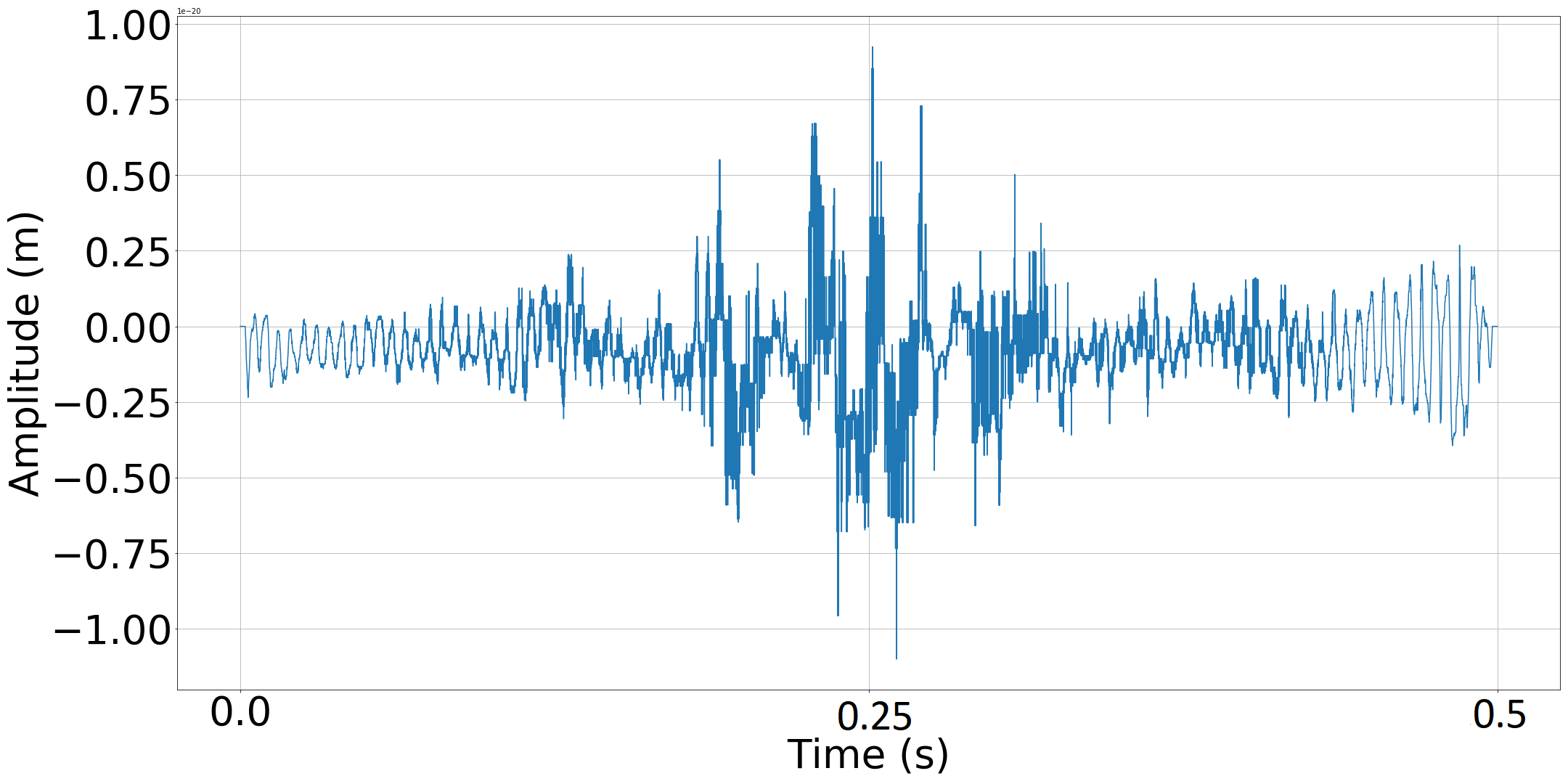}
            \includegraphics[width=6cm,height=4cm]{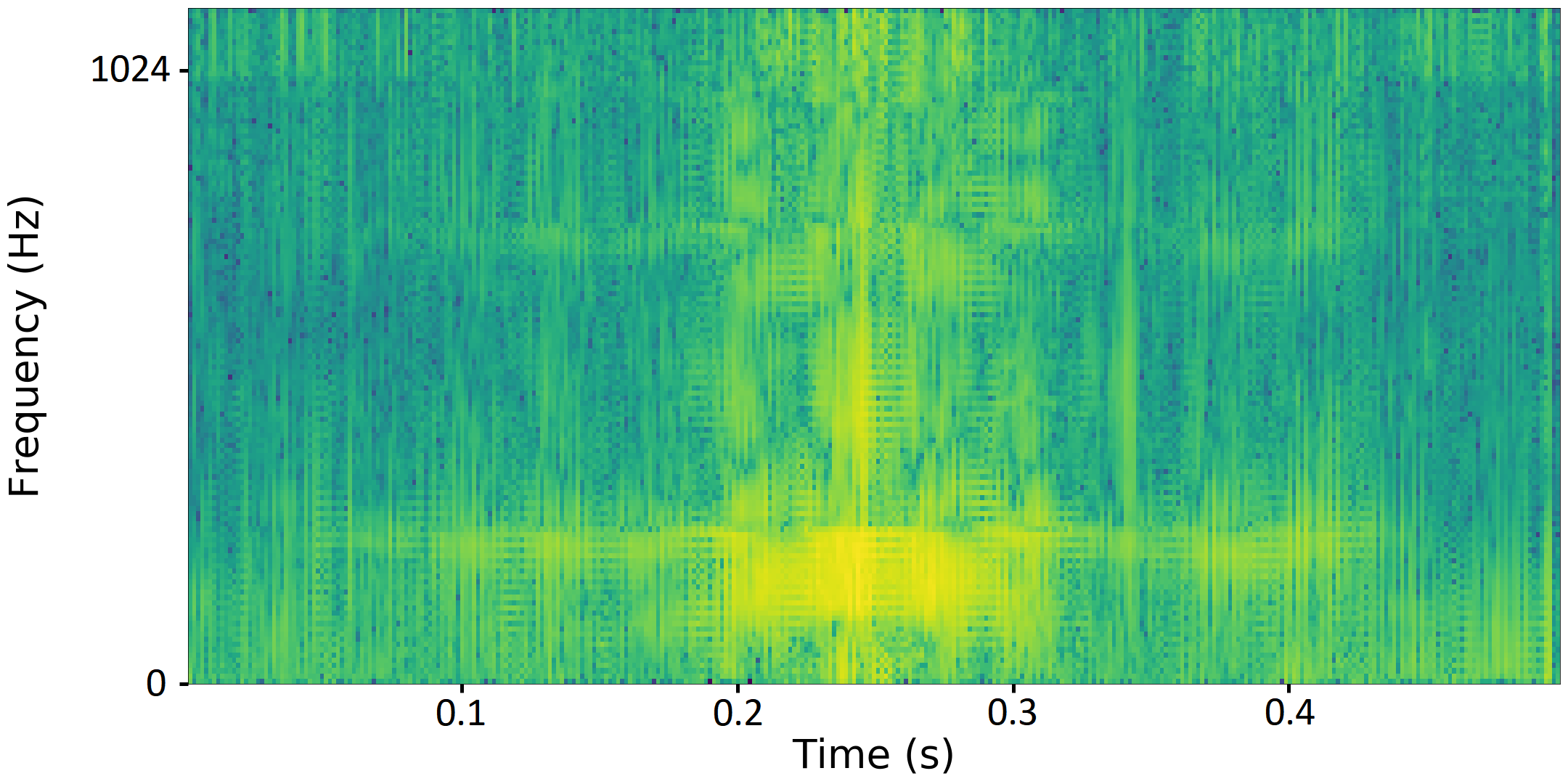}
          \includegraphics[width=5.5cm,height=4cm]{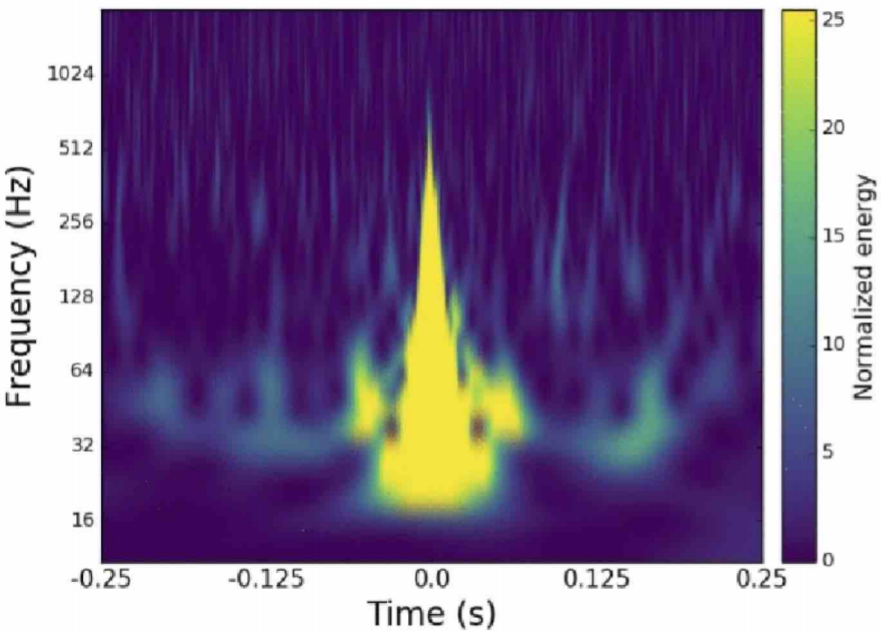}
\end{subfigure}

\begin{subfigure}[]{}
            \includegraphics[width=6cm,height=4cm]{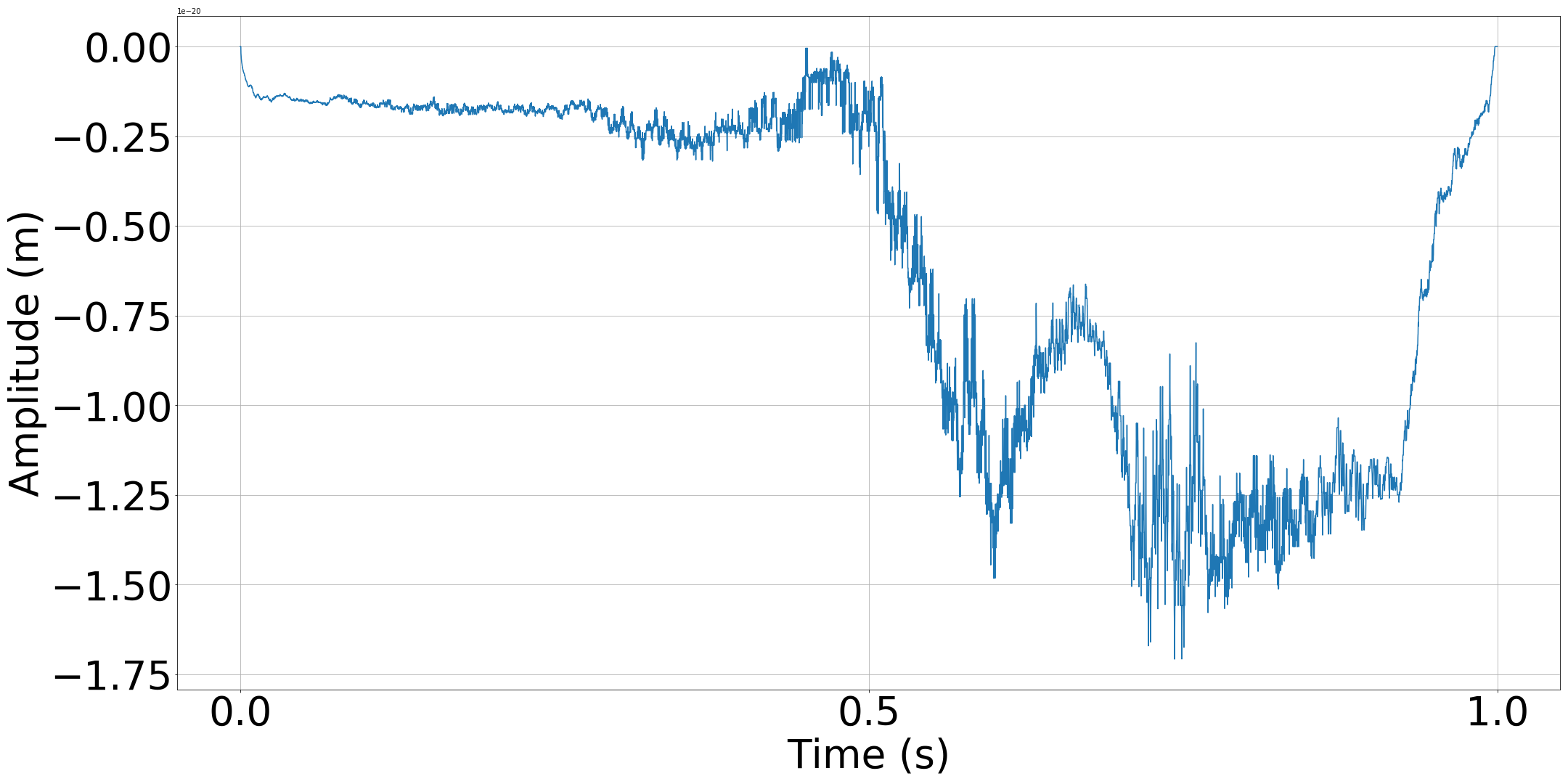}
            \includegraphics[width=6cm,height=4cm]{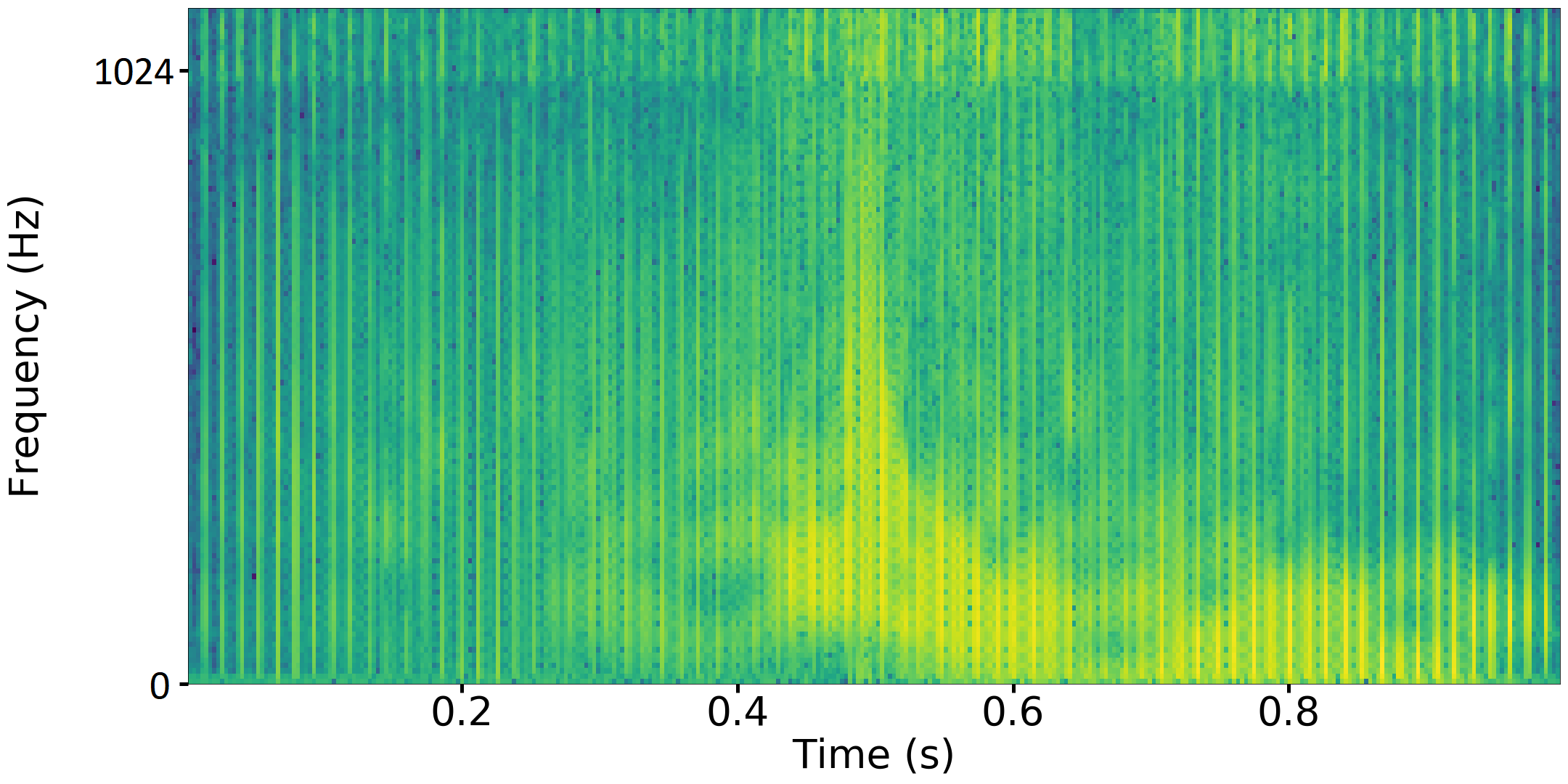}
          \includegraphics[width=5.5cm,height=4cm]{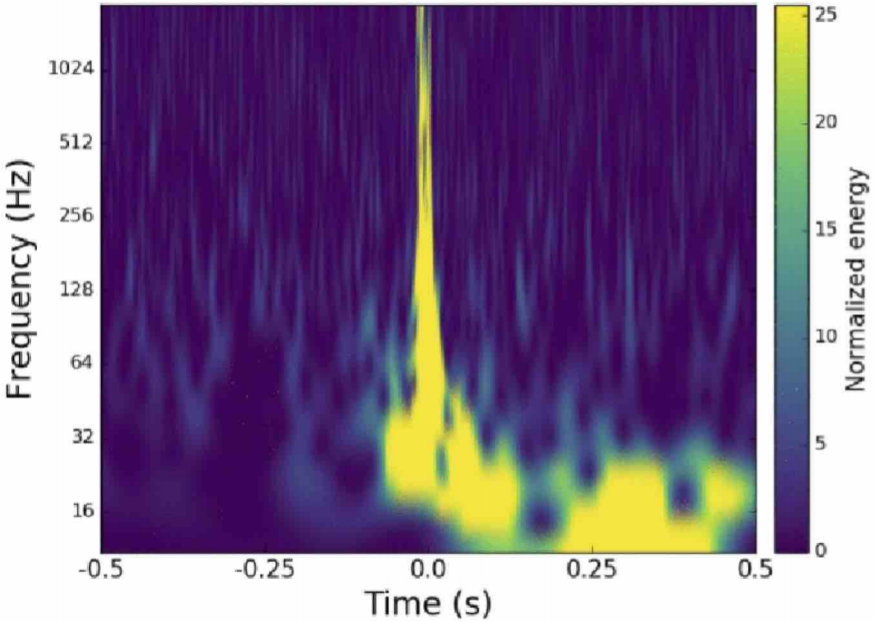}
\end{subfigure}

\begin{subfigure}[]{}
            \includegraphics[width=6cm,height=4cm]{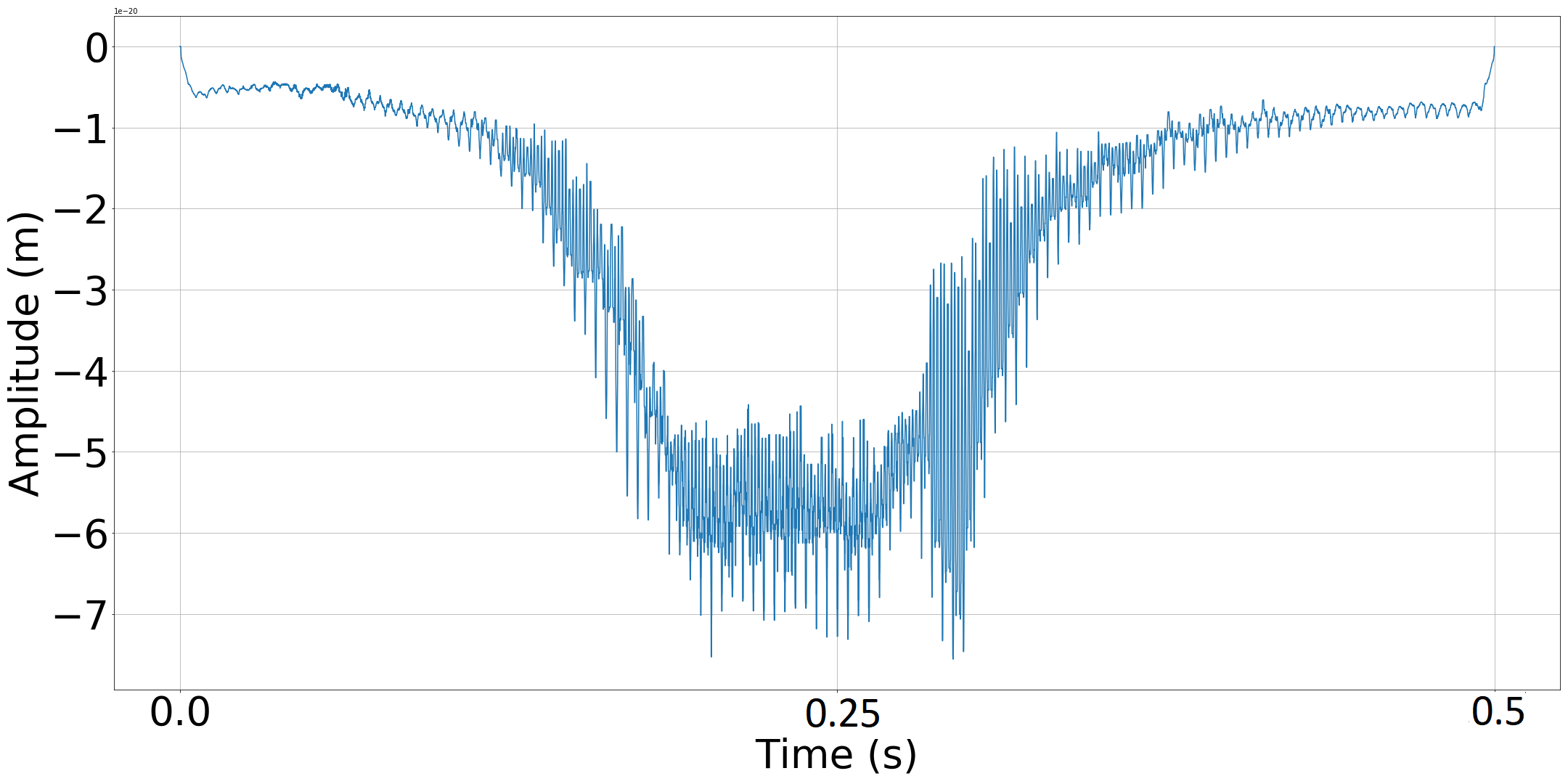}
            \includegraphics[width=6cm,height=4cm]{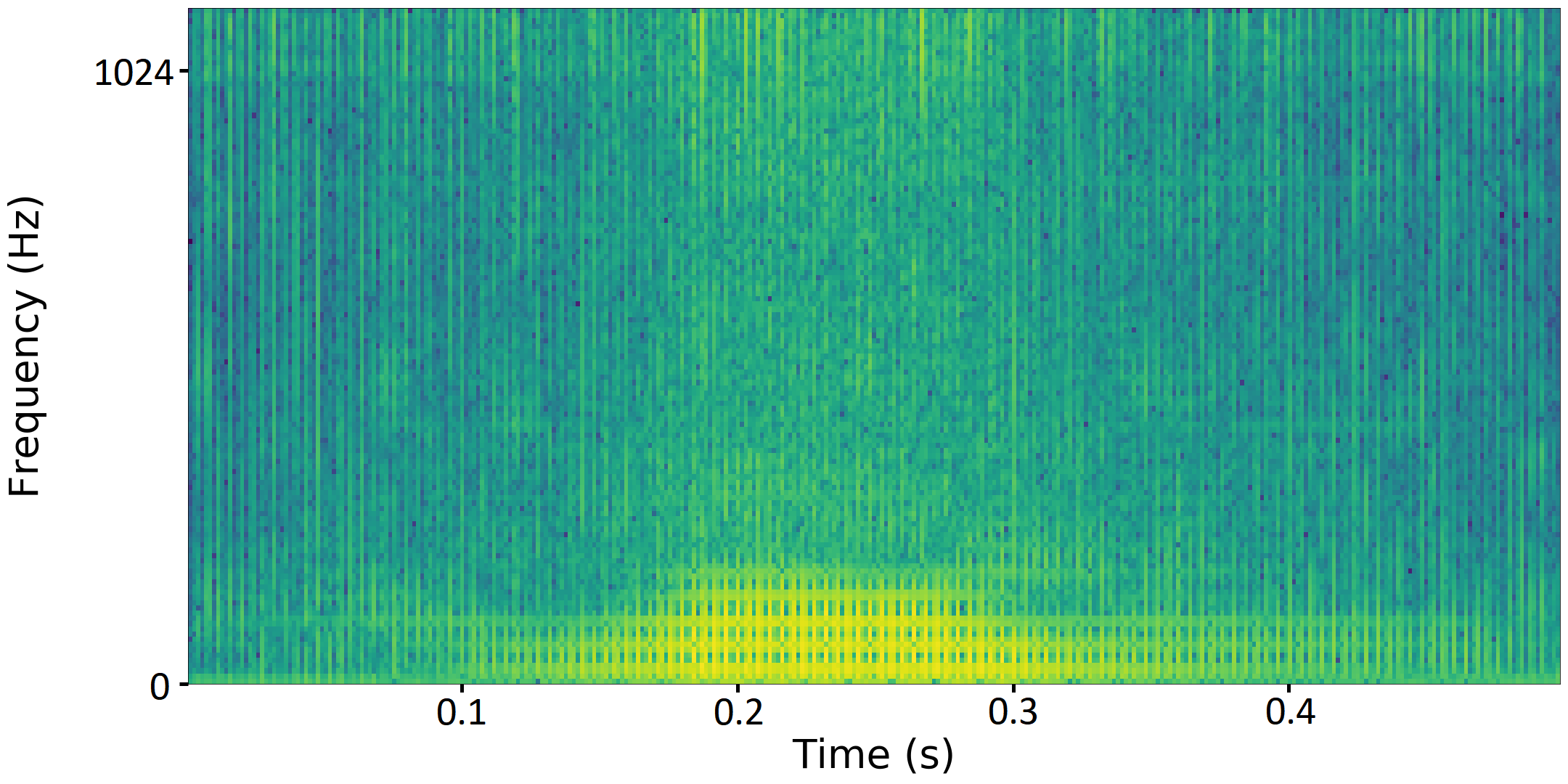}
          \includegraphics[width=5.5cm,height=4cm]{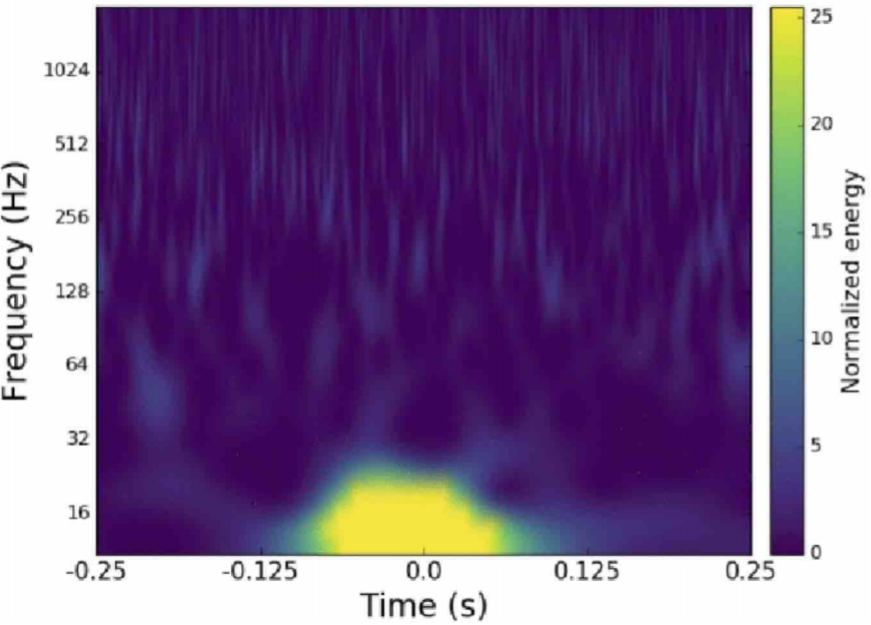}
\end{subfigure}

\begin{subfigure}[]{}
            \includegraphics[width=6cm,height=4cm]{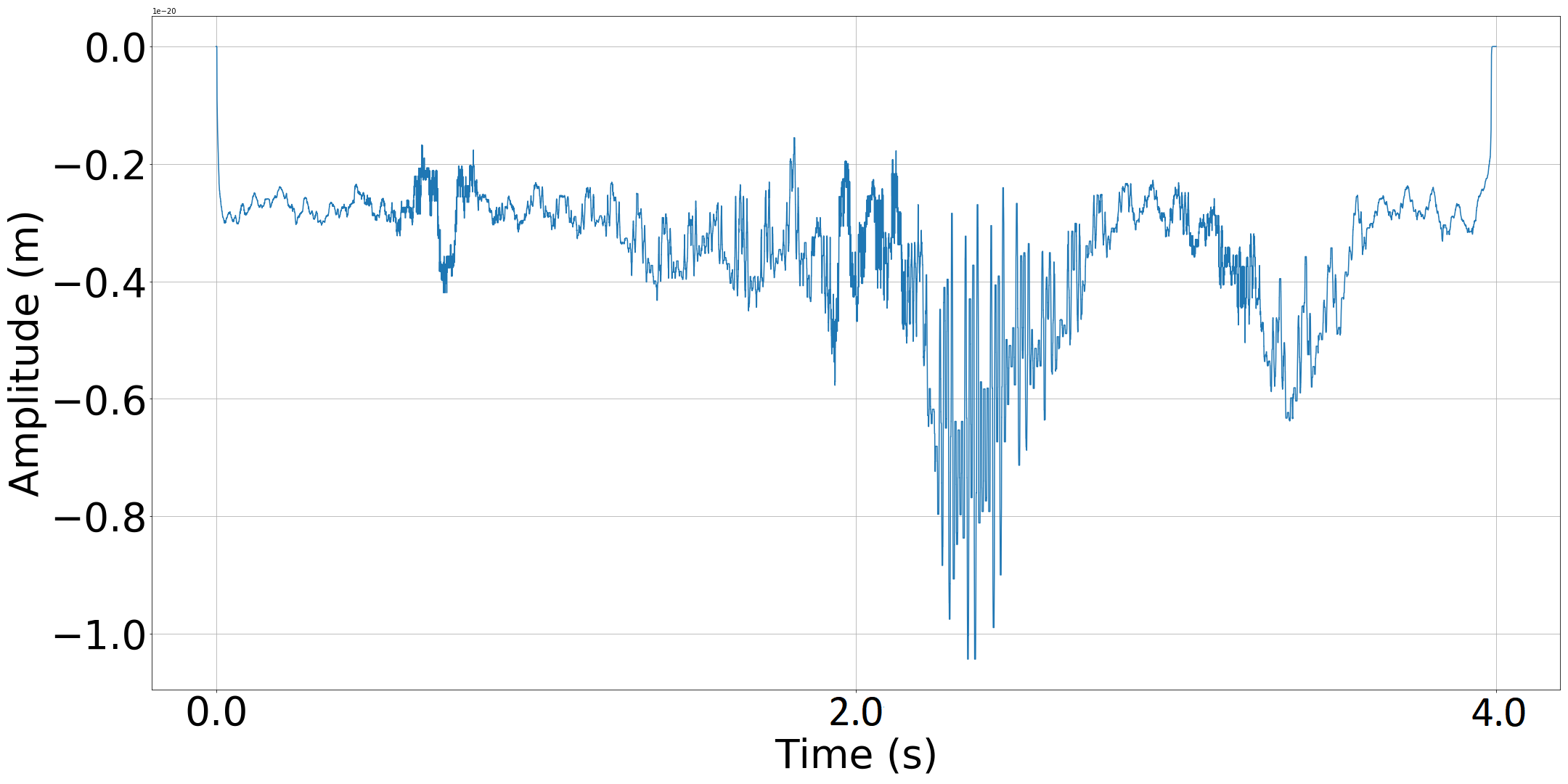}
            \includegraphics[width=6cm,height=4cm]{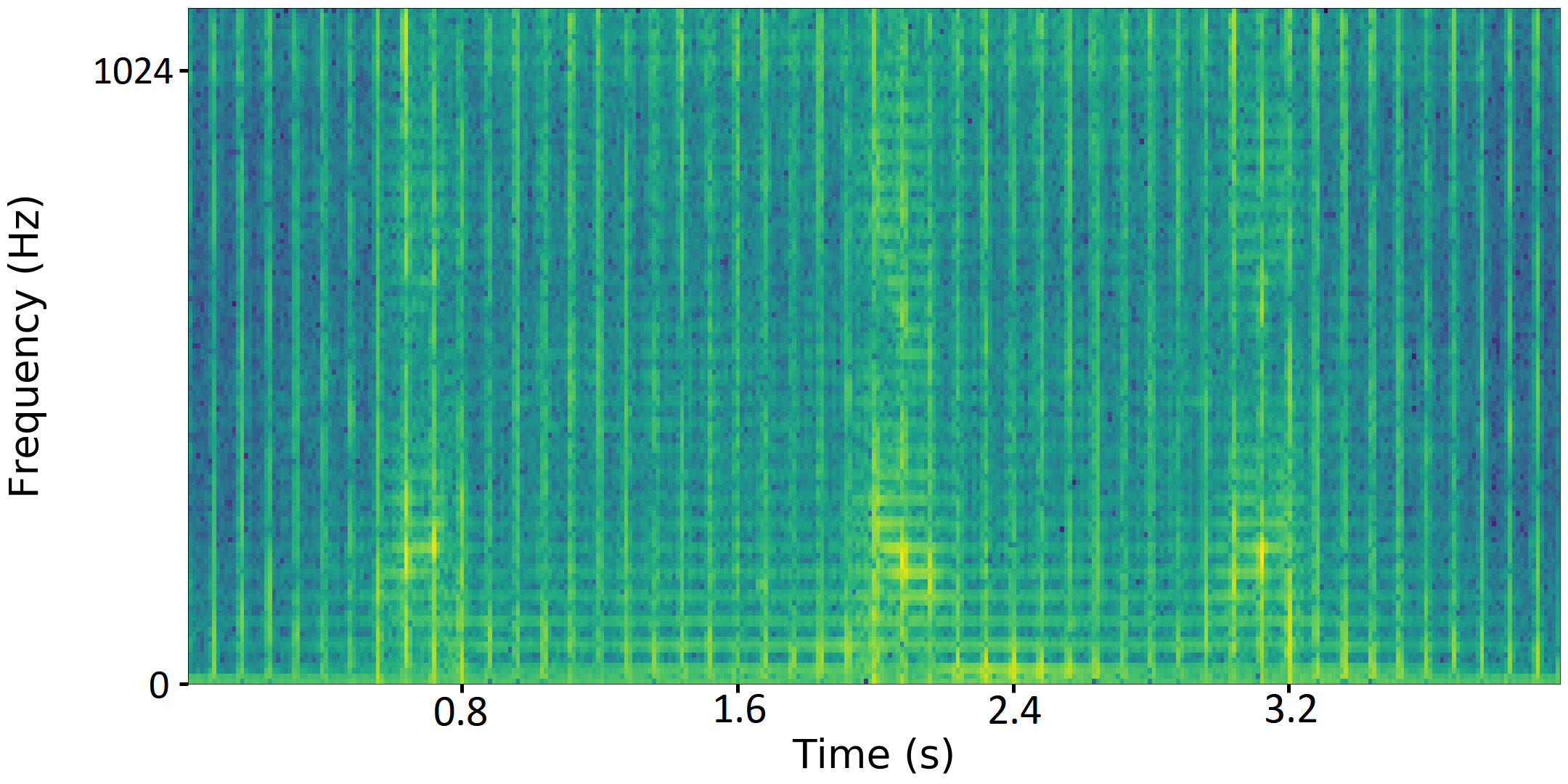}
          \includegraphics[width=5.5cm,height=4cm]{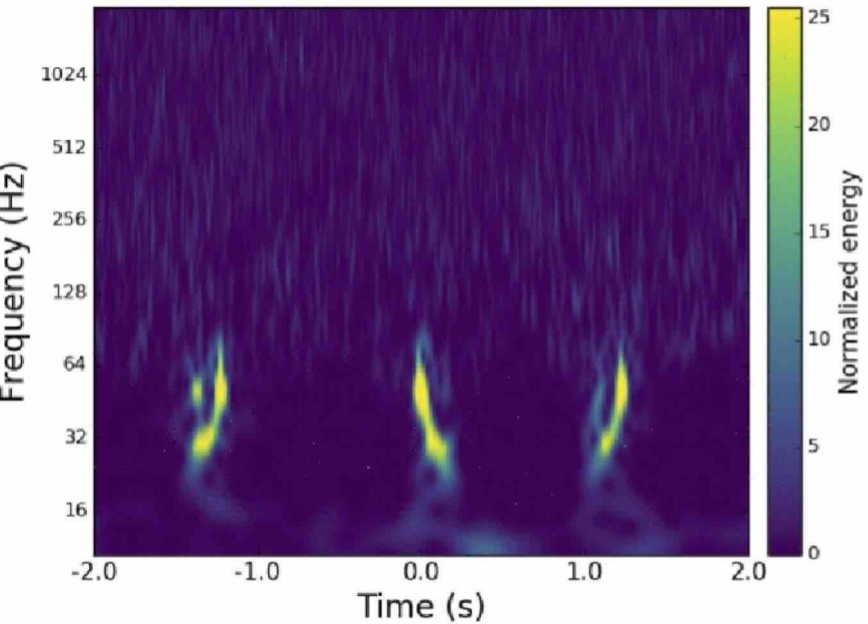}
\end{subfigure}

\begin{subfigure}[]{}
            \includegraphics[width=6cm,height=4cm]{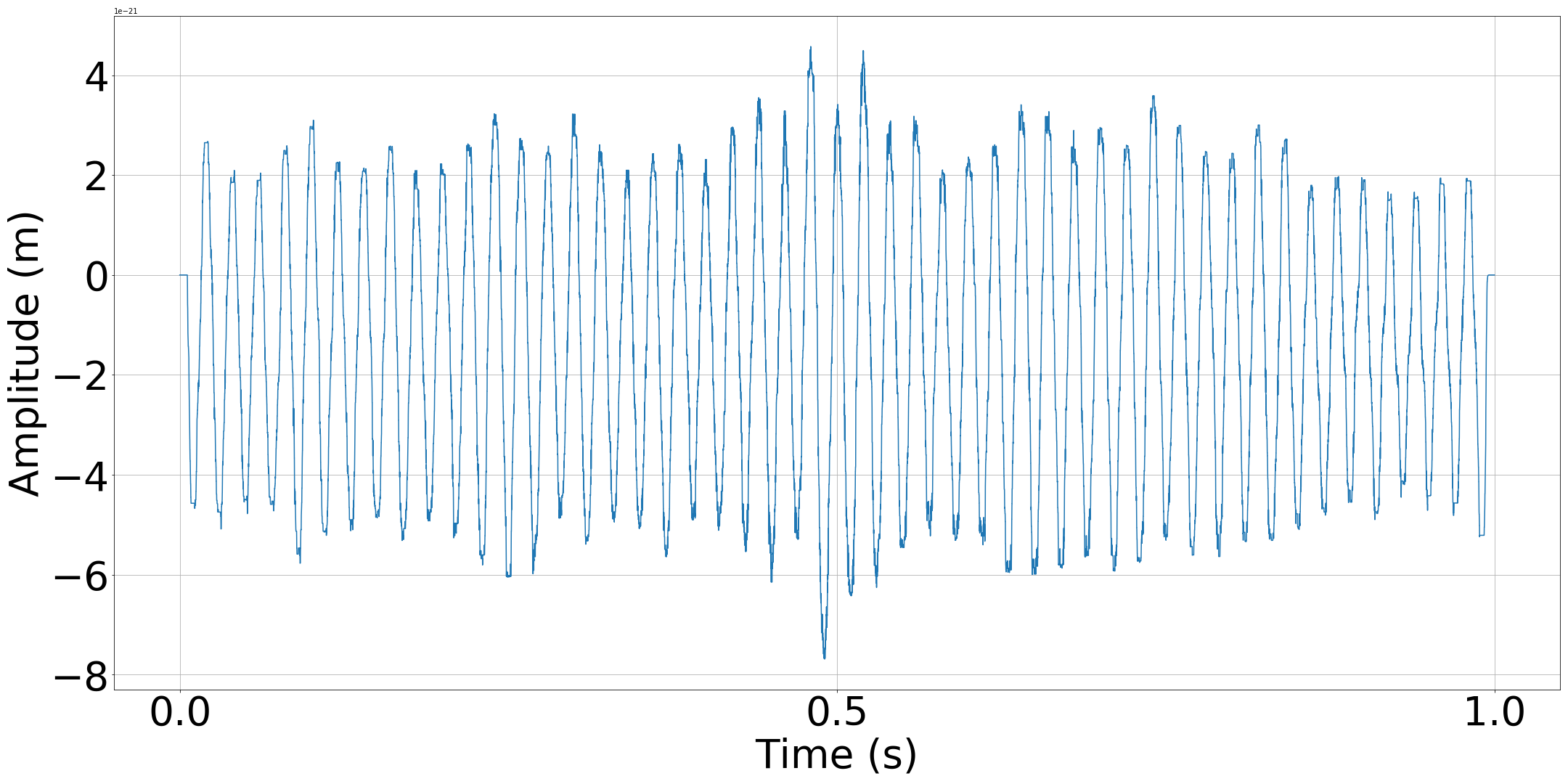}
            \includegraphics[width=6cm,height=4cm]{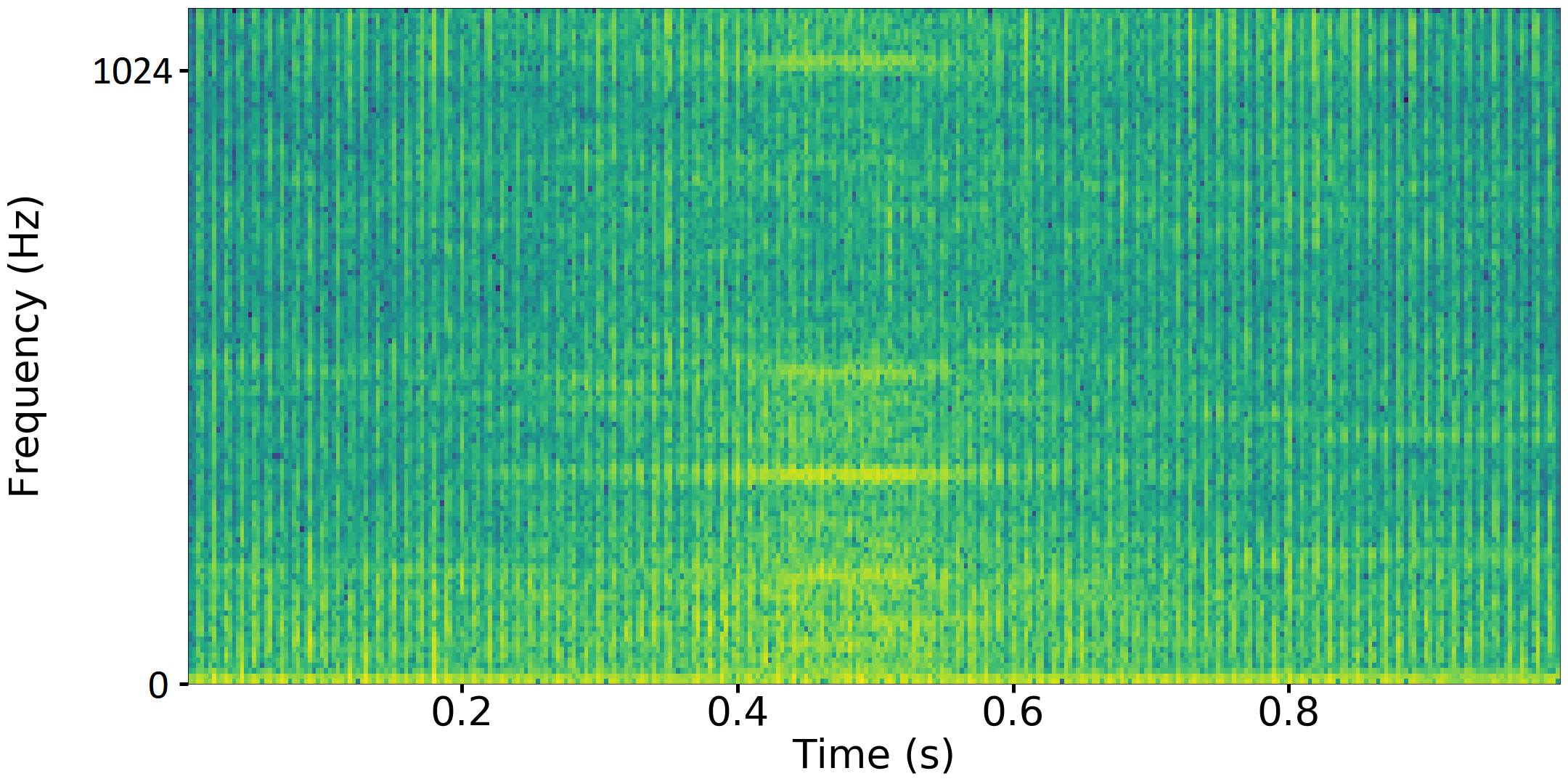}
          \includegraphics[width=5.5cm,height=4cm]{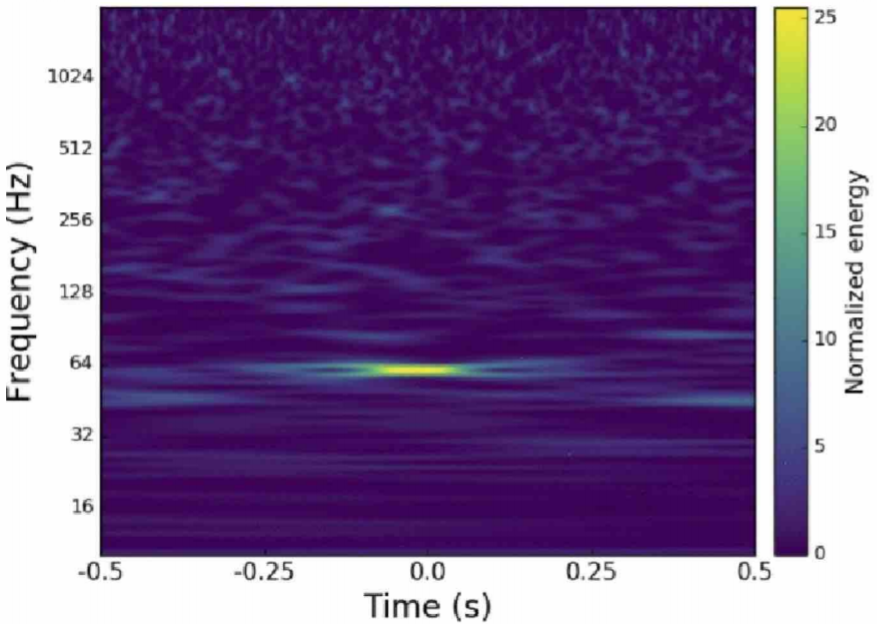}
\end{subfigure}

\end{figure*}

\begin{figure*}

\begin{subfigure}[]{}
            \includegraphics[width=6cm,height=4cm]{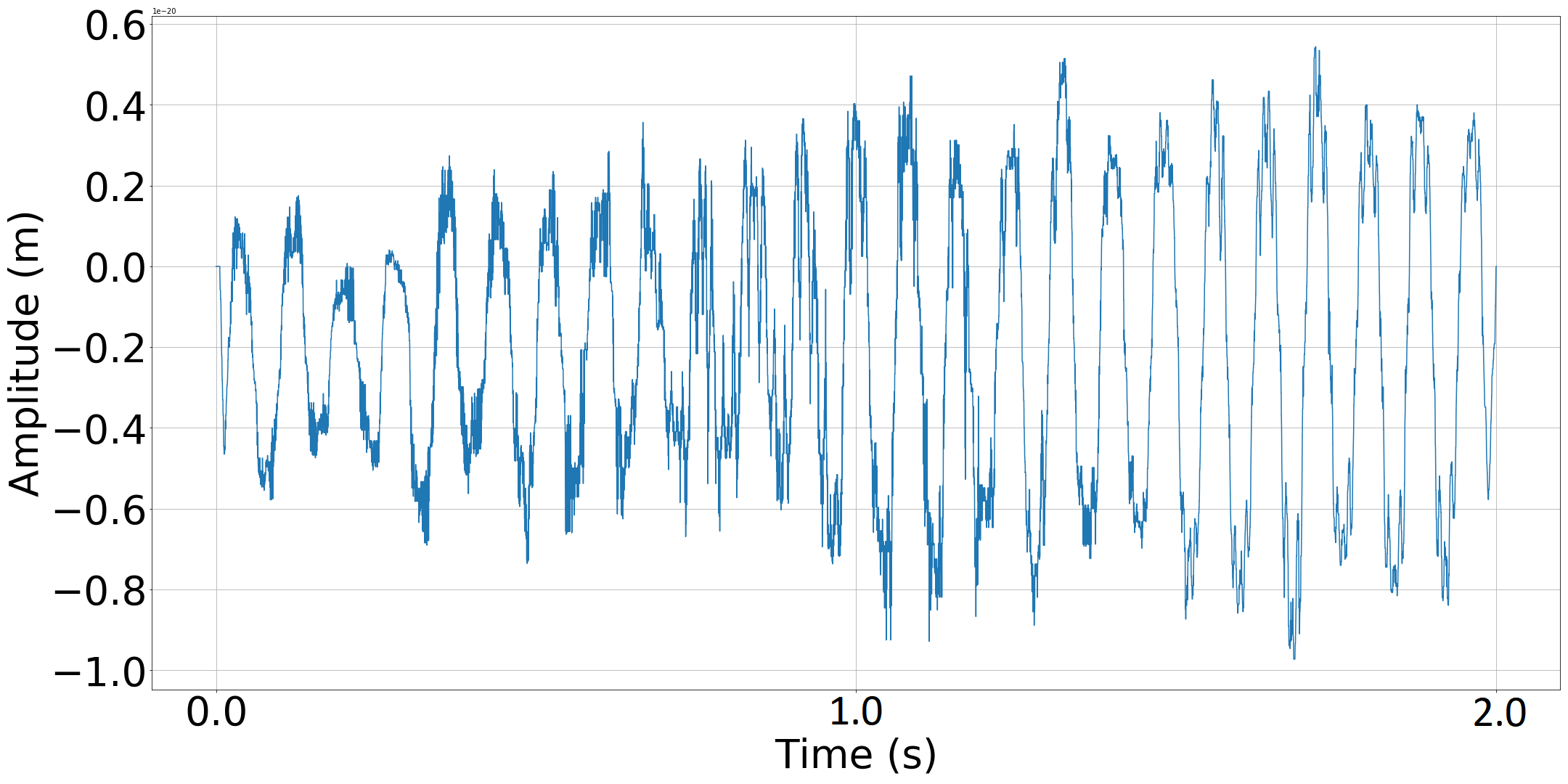}
            \includegraphics[width=6cm,height=4cm]{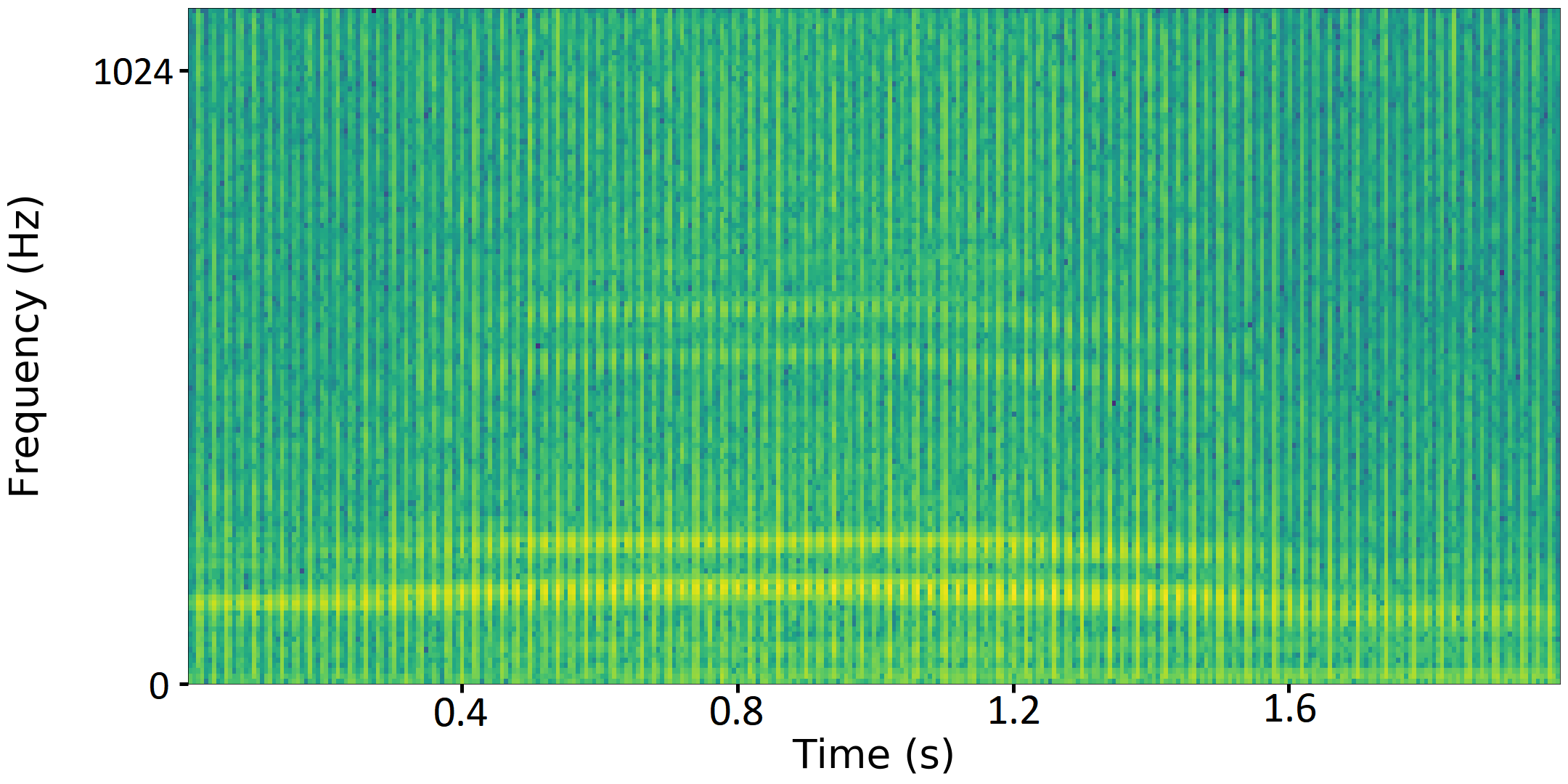}
          \includegraphics[width=5.5cm,height=4cm]{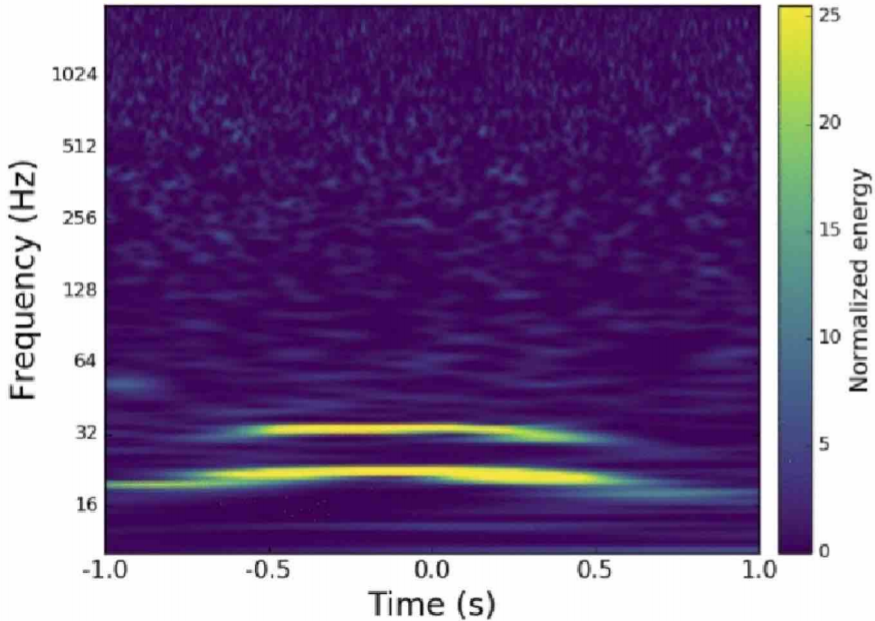}
\end{subfigure}

\begin{subfigure}[]{}
            \includegraphics[width=6cm,height=4cm]{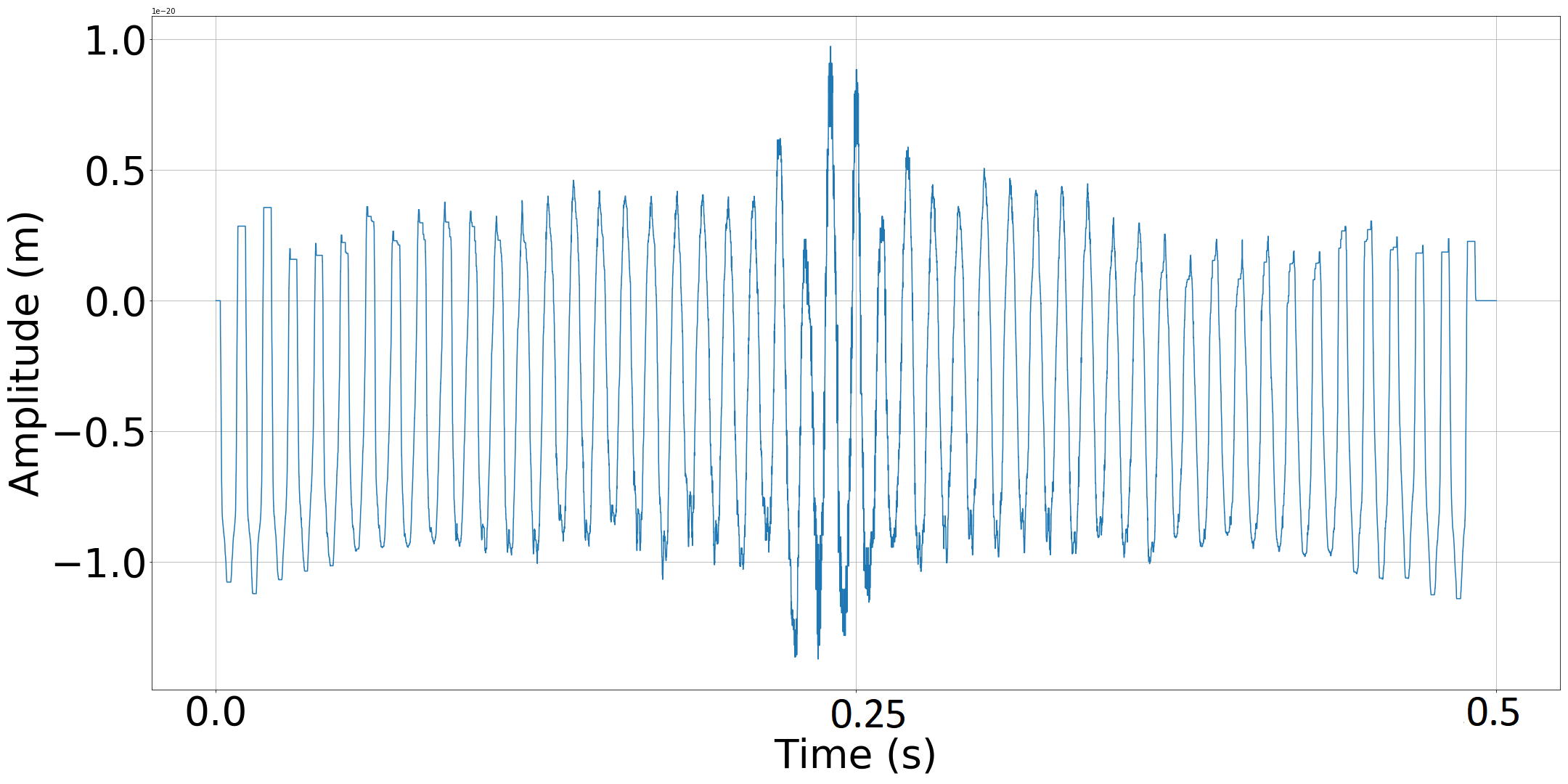}
            \includegraphics[width=6cm,height=4cm]{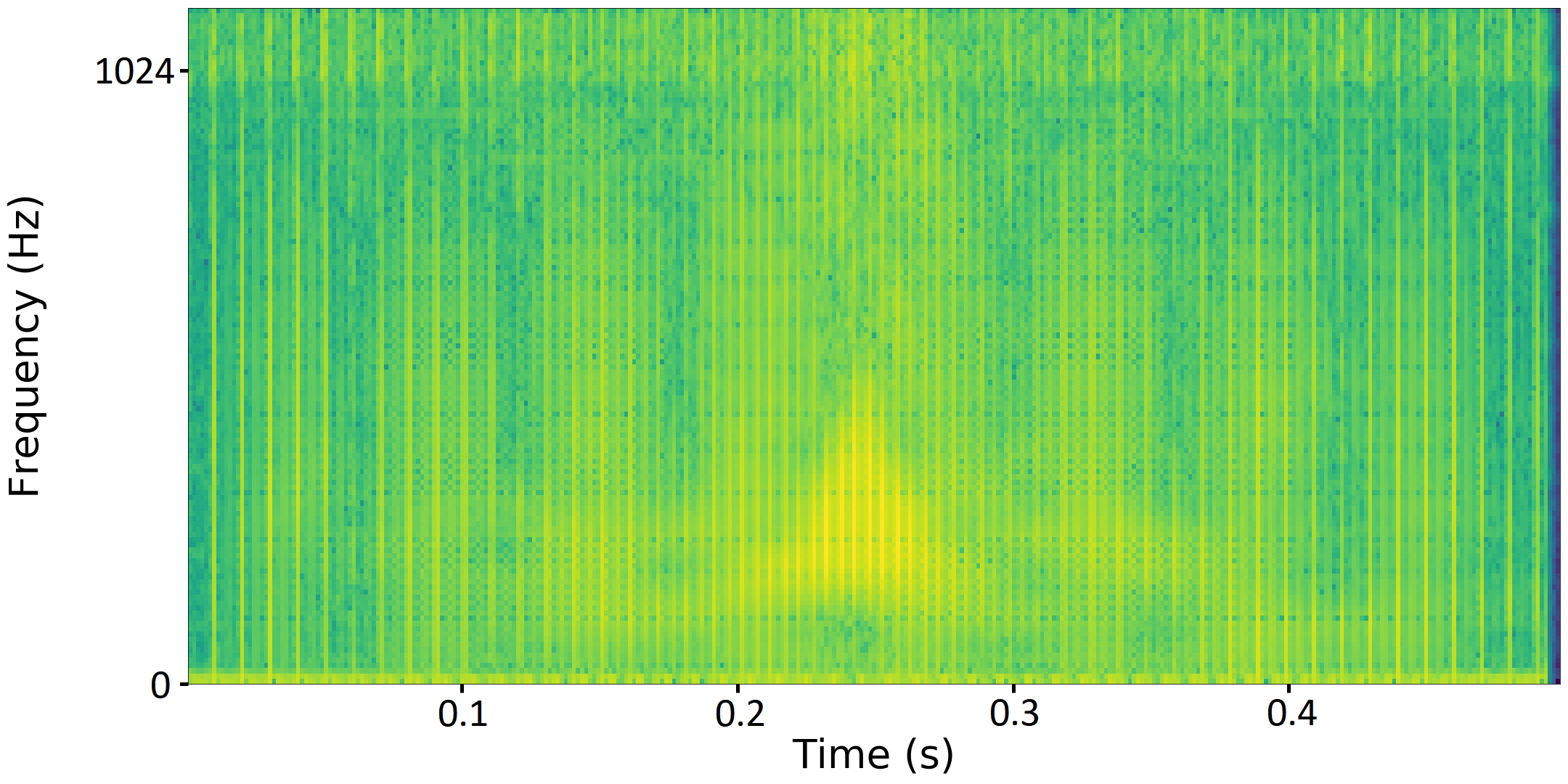}
          \includegraphics[width=5.5cm,height=4cm]{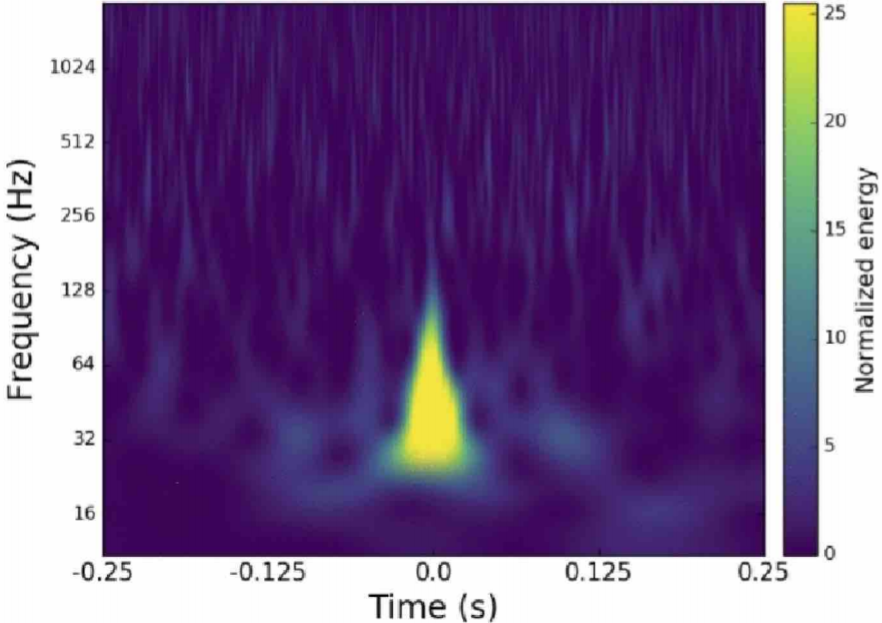}
\end{subfigure}

\begin{subfigure}[]{}
            \includegraphics[width=6cm,height=4cm]{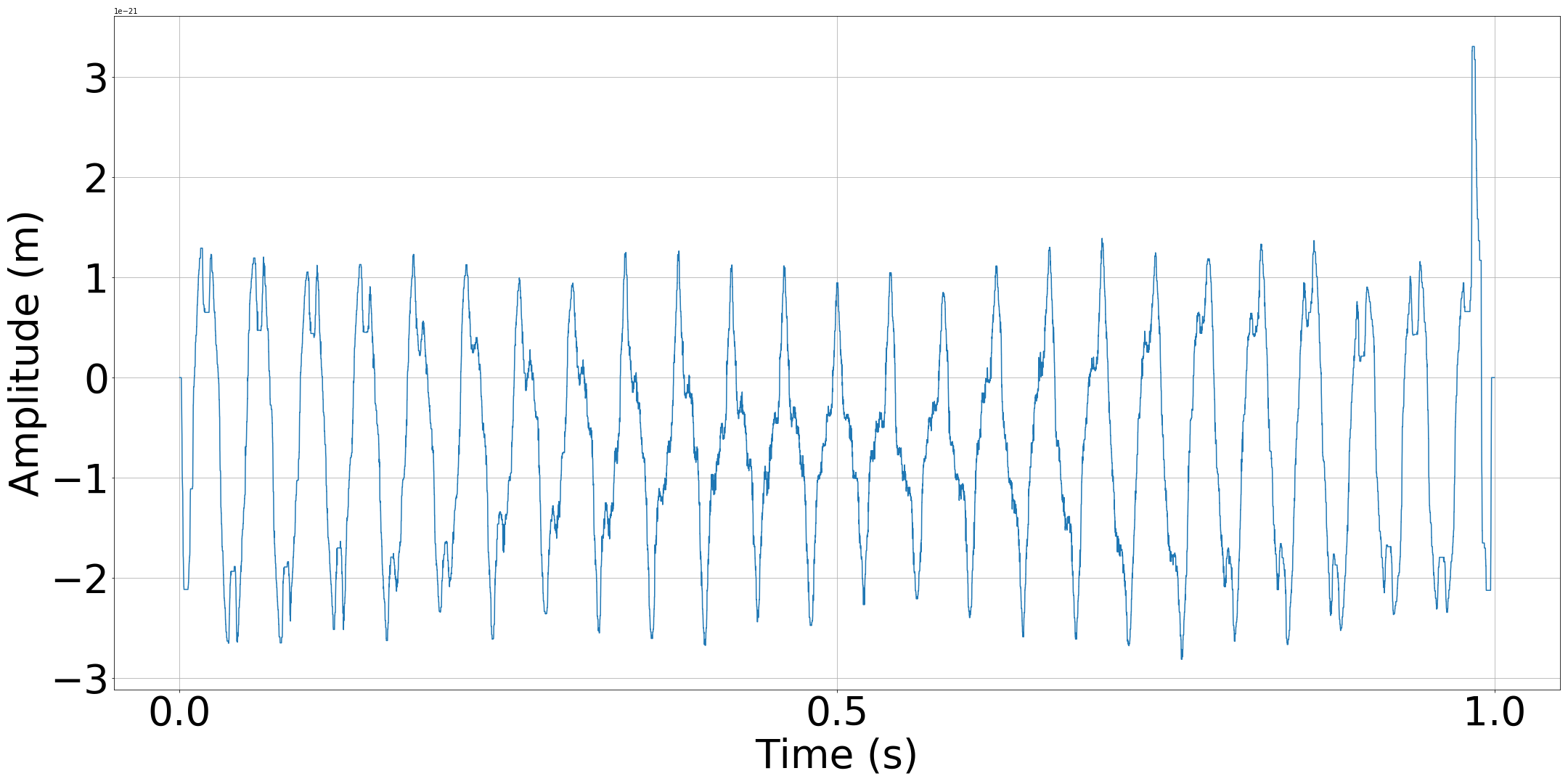}
            \includegraphics[width=6cm,height=4cm]{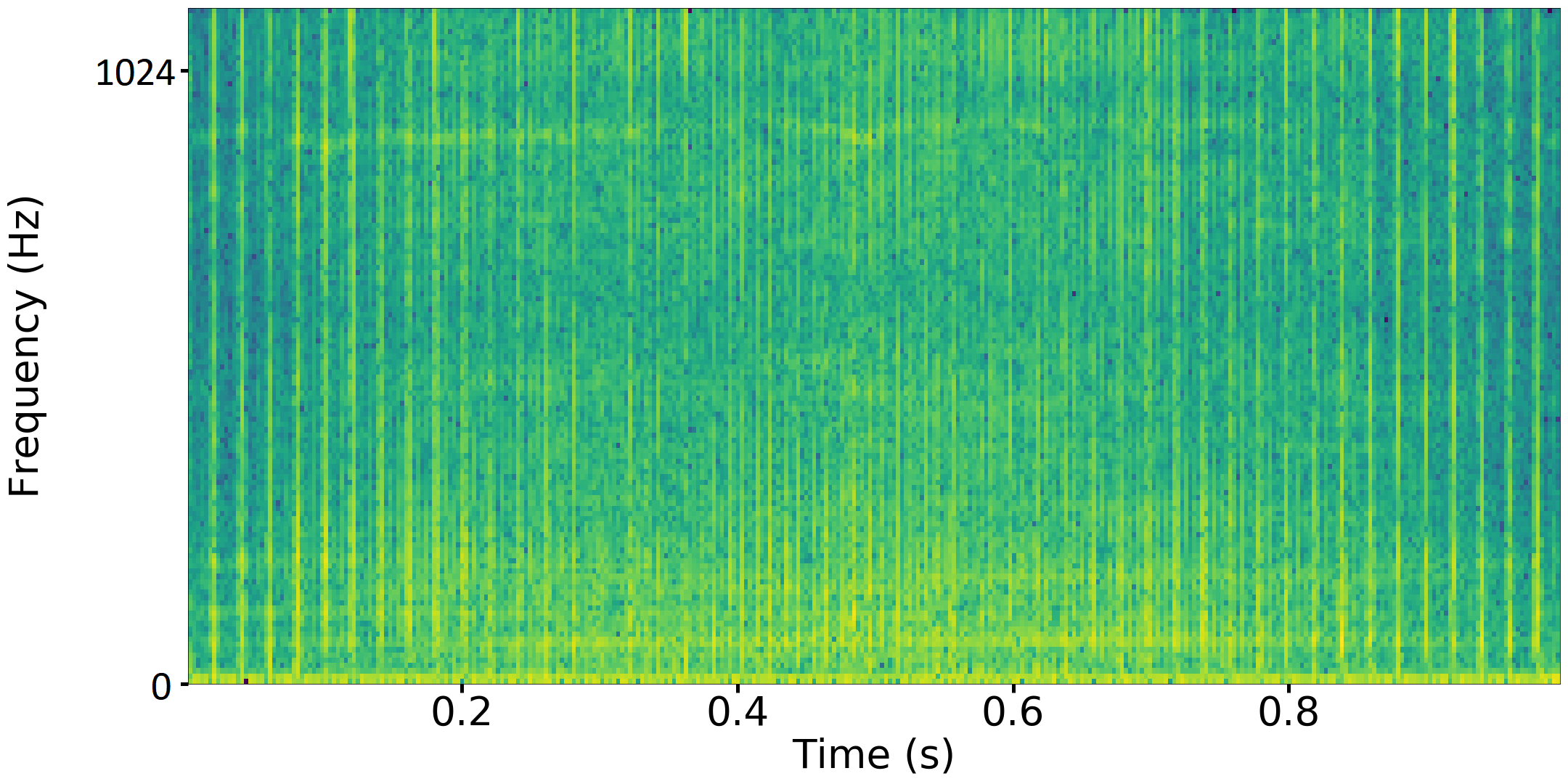}
          \includegraphics[width=5.5cm,height=4cm]{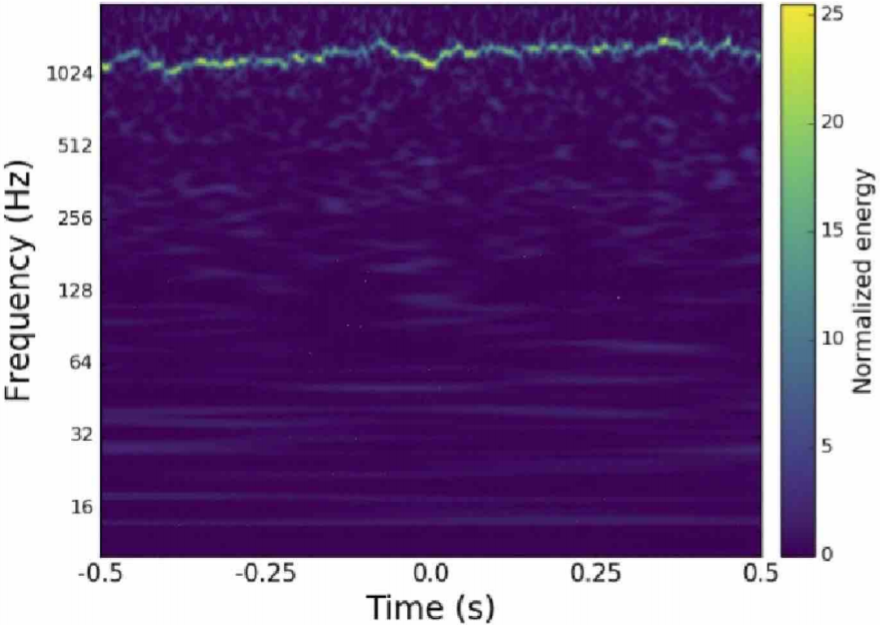}
\end{subfigure}

\begin{subfigure}[]{}
            \includegraphics[width=6cm,height=4cm]{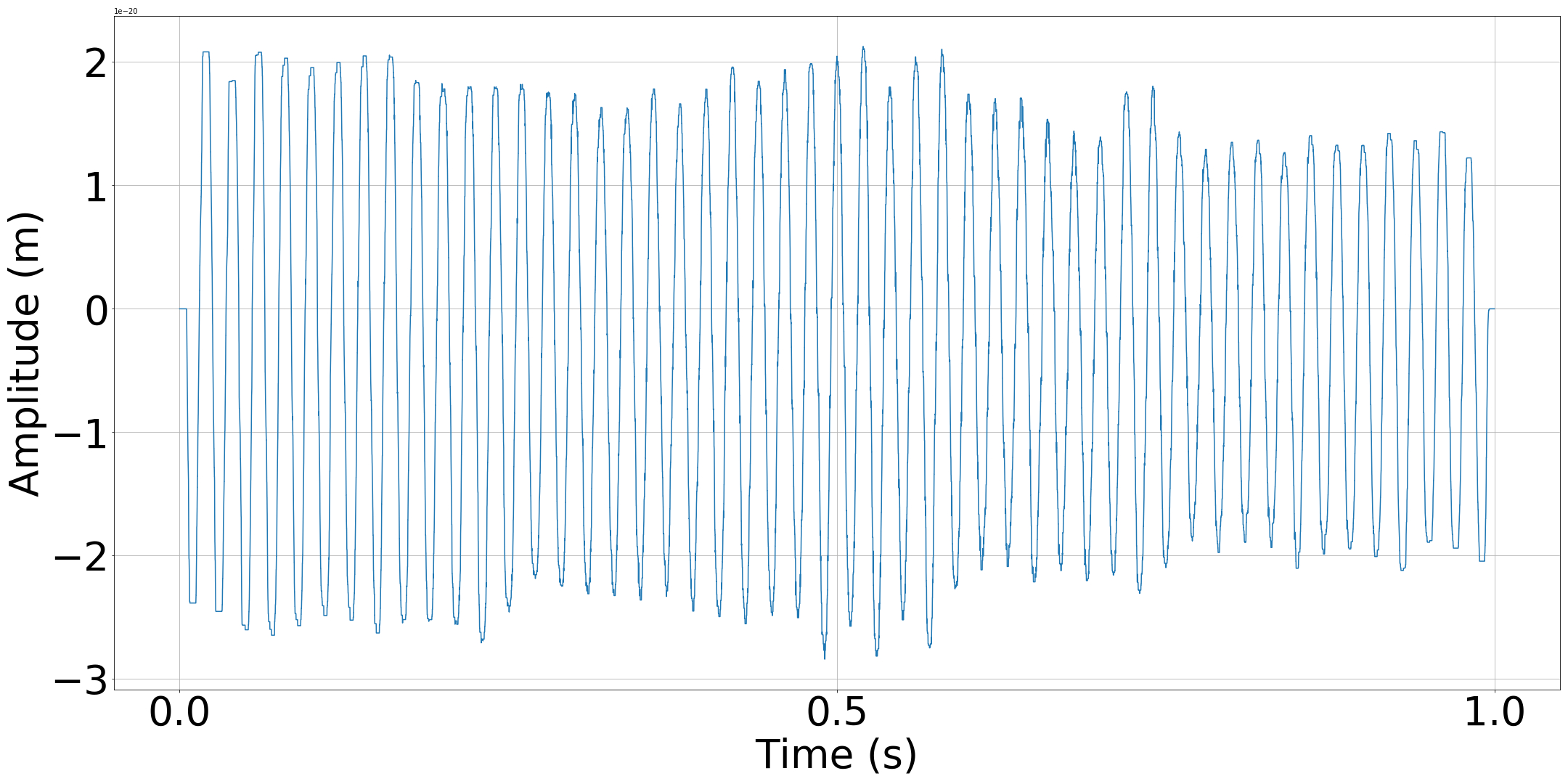}
            \includegraphics[width=6cm,height=4cm]{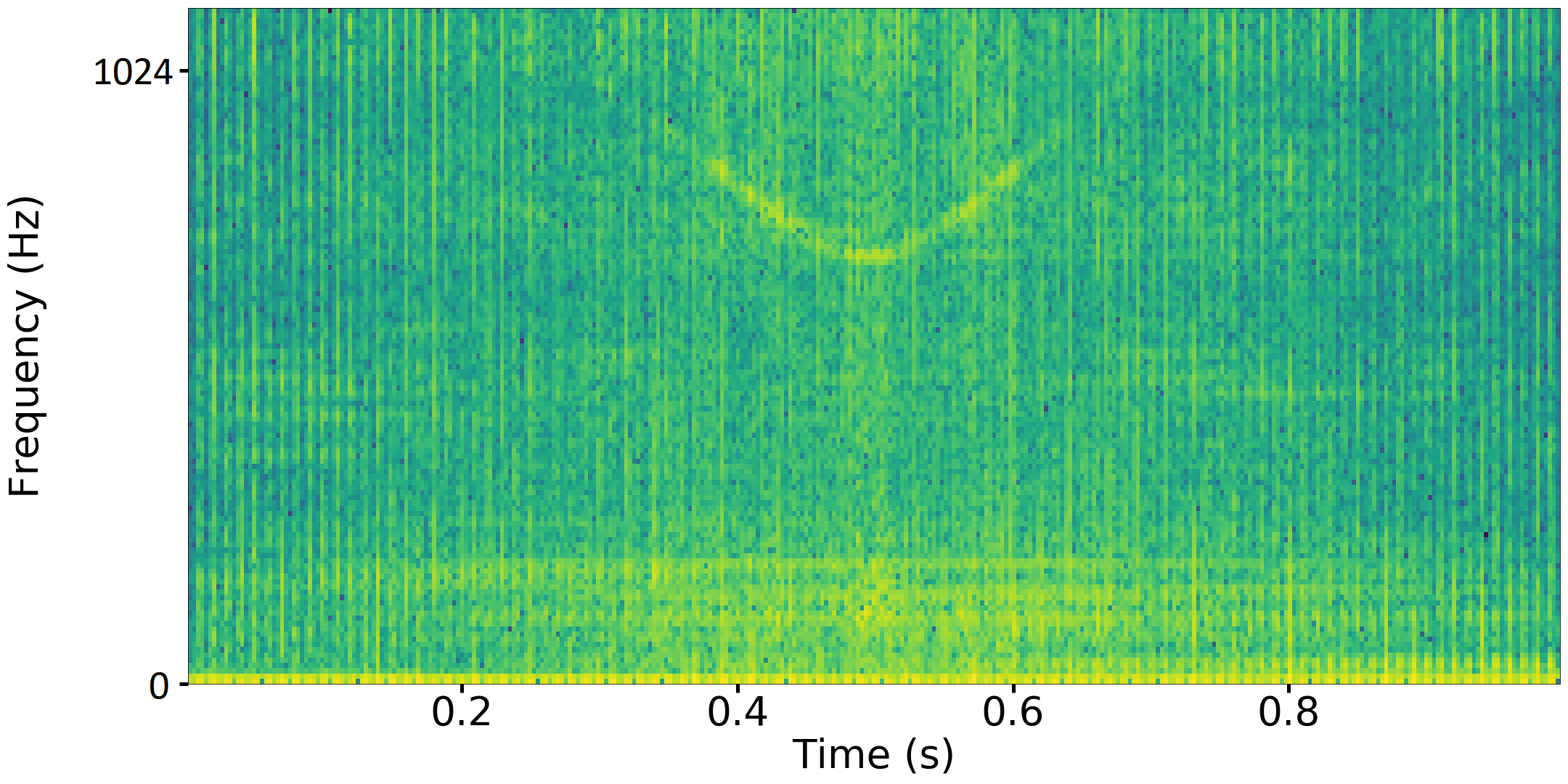}
          \includegraphics[width=5.5cm,height=4cm]{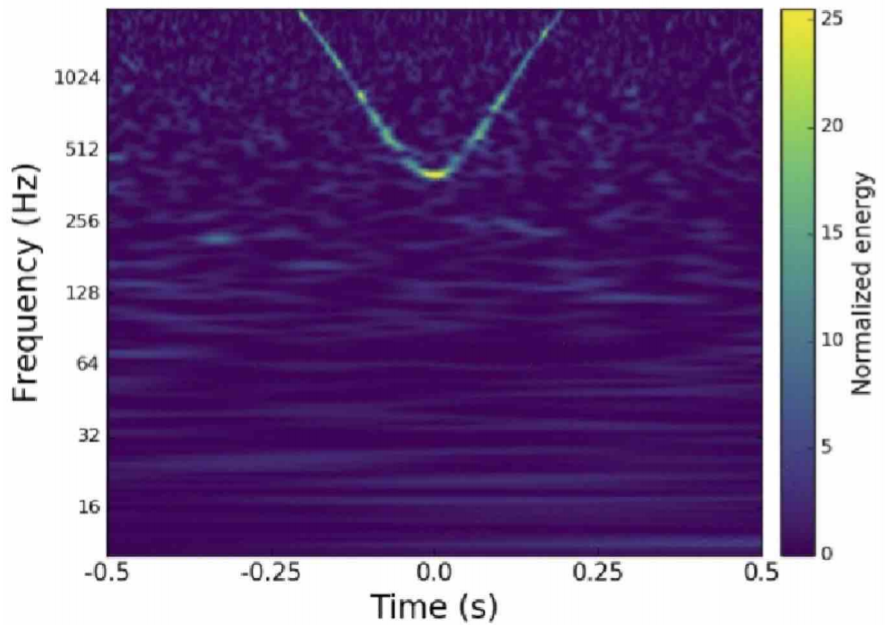}
\end{subfigure}
\end{figure*}

\begin{figure*}
\begin{subfigure}[]{}
            \includegraphics[width=6cm,height=4cm]{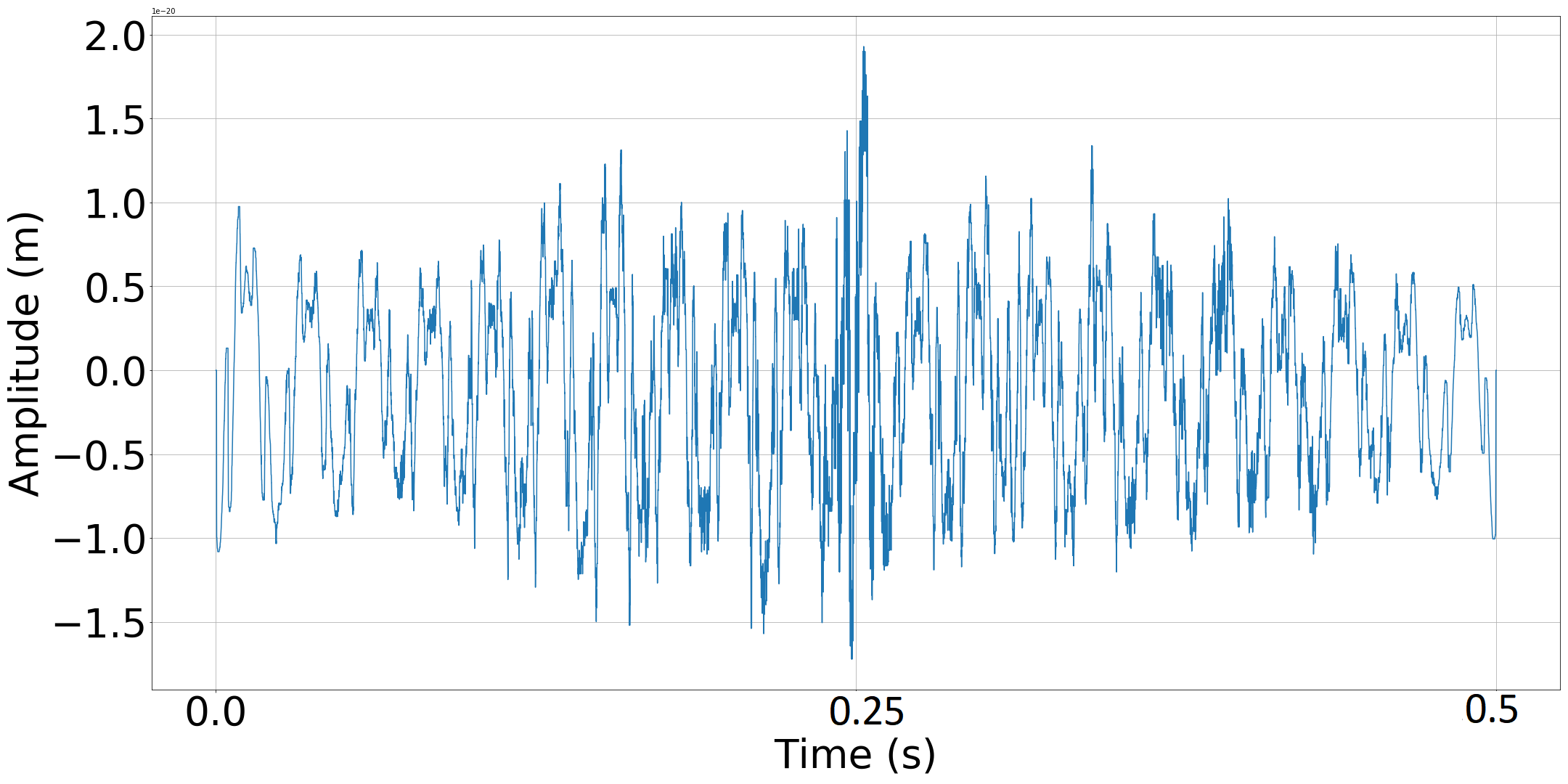}
            \includegraphics[width=6cm,height=4cm]{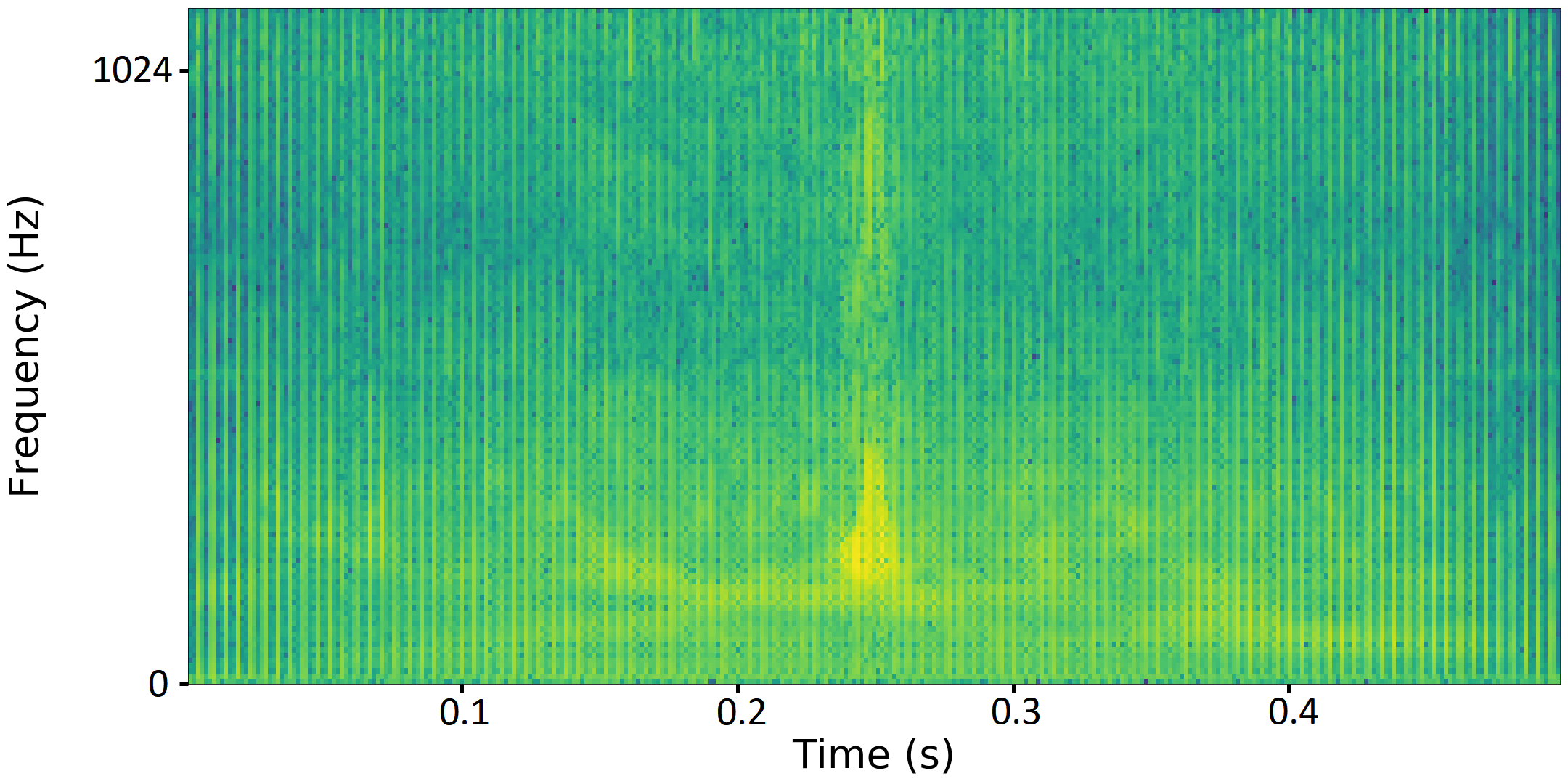}
          \includegraphics[width=5.5cm,height=4cm]{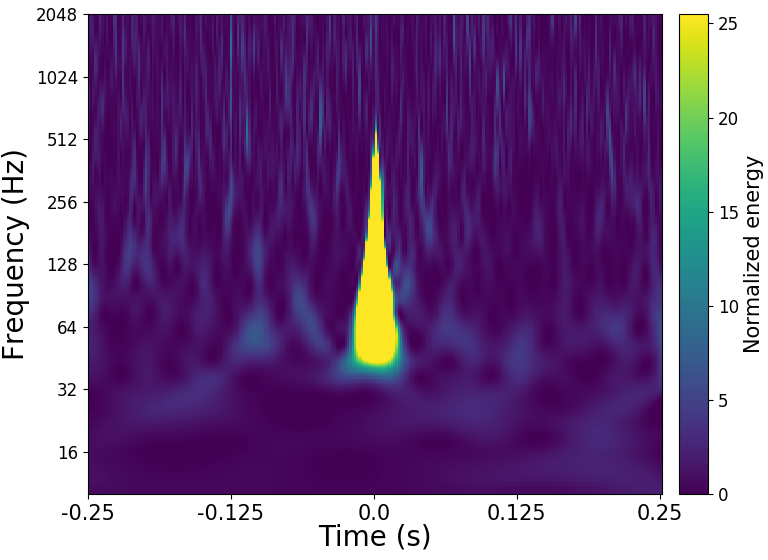}
\end{subfigure}
\caption{Waveform of the various glitches obtained from \texttt{Image to Sound} method. Rightmost image represents the timeseries signal, middle image represent the spectrogram generated from \texttt{Image to Sound} method and the righmost image represents the original image that was used for the method. The glitches are arranged in sequence from top to bottom as: (a) koi fish, (b) light modulation, (c) low frequency burst, (d) paired dove, (e) power line, (f) scattered light, (g) Tomte, (h) wandering line, (i) whistle, (j) Blip. Blip glitch has not been used but is used as a for validating the waveform \citep{cabero2019blip}.}
\label{fig8}
\end{figure*}

\begin{figure*}
\centering
\begin{subfigure}[Waveform for CCSNe model A1B1G1$^{R}$]{\includegraphics[width=8cm,height=6cm]{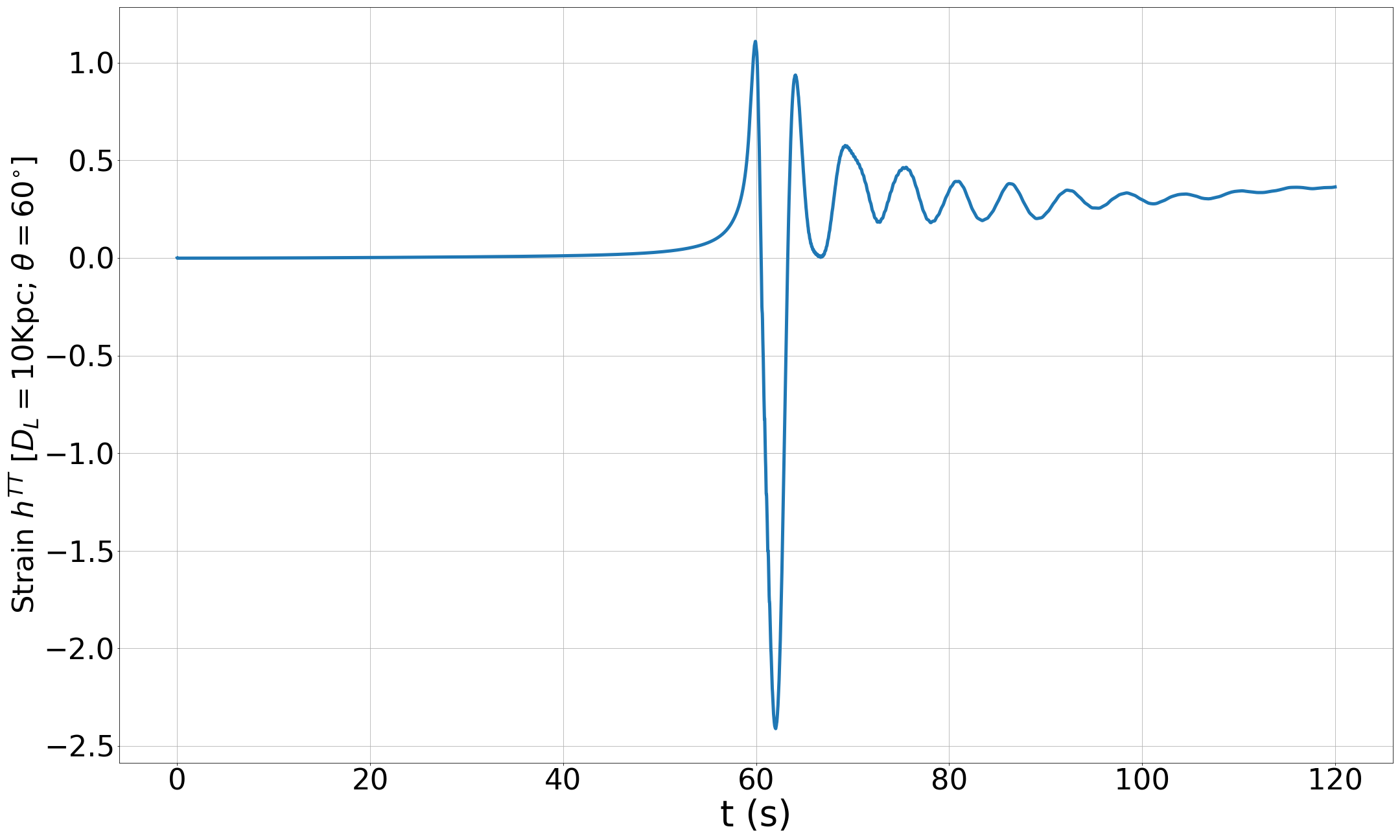}}\end{subfigure}
\begin{subfigure}[Waveform for CCSNe model A3B1G1$^{R}$]{\includegraphics[width=8cm,height=6cm]{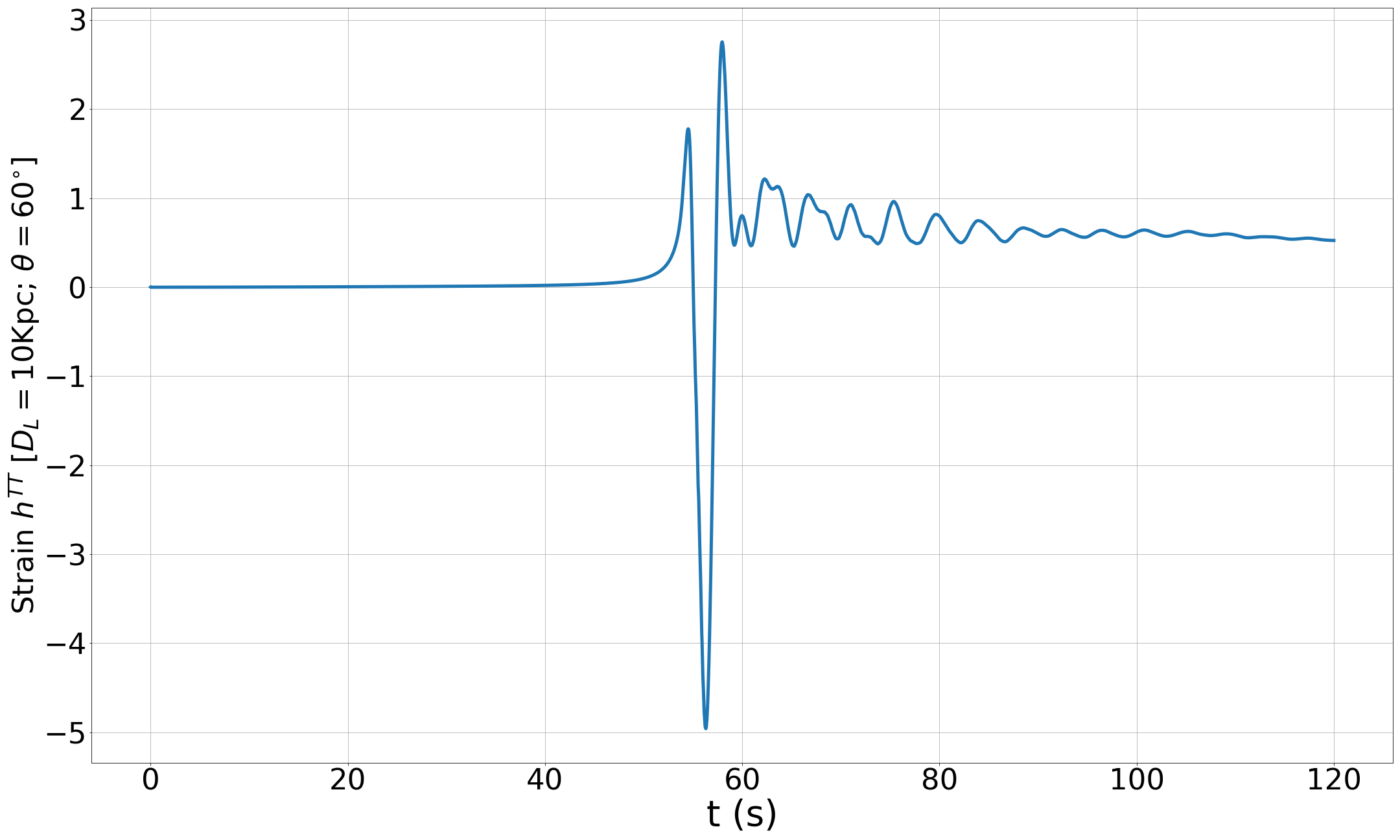}}\end{subfigure}
\caption{}
\label{fig9}
\end{figure*}

\bsp	% typesetting comment
\label{lastpage}
\end{document}